\documentclass[english]{article}
\usepackage[T1]{fontenc}
\usepackage[latin9]{inputenc}
\usepackage{geometry}
\usepackage{color}
\usepackage{calc}
\usepackage{amsmath}
\usepackage{amssymb}
\usepackage{graphicx}

\makeatletter

\newcommand{\lyxdot}{.}

\newcommand{\lyxaddress}[1]{
	\par {\raggedright #1
	\vspace{1.4em}
	\noindent\par}
}

\makeatother

\usepackage{babel}
\begin{document}
\title{Resolving Collisions for the Gipps Car-Following Model}
\author{Leonhard L\"ucken$^{1}$}

\maketitle

\lyxaddress{$^{1}$German Aerospace Center (DLR), Institute of Transportation
Systems, Rutherfordstraße 2, 12489 Berlin, Germany}

\begin{abstract}
The Gipps car-following model is a widely used tool for studying and simulation traffic dynamics.
Despite its popularity an often disregarded property is that under heterogeneous parametrization
on the individual vehicles in the traffic flow, the model may produce collisions. This stands in
crude contrast to the principle, from which the model was derived: drive as fast as possible while
guaranteeing a safe headway in case that the leading vehicle starts braking hard. Indeed, Gipps
proof for the model being collision-free only holds for ensembles of identical vehicles.

In this work we examine the circumstances leading to collisions in heterogeneous ensembles and propose
a natural model extension, which conveys the original models principles to situations, where 
collisions occur. For these cases we present analytical and numerical results on the stability 
of the equilibrium flow.

\end{abstract}

\section{Introduction}

Numerical simulation is an important tool for the understanding of
dynamical phenomena arising in traffic systems as, for instance, traffic
jam formation and propagation. Traffic simulations can also help to
predict the performance of specific road network elements or their
configurations, e.g., a novel traffic light algorithm. They are valuable
tools for the guidance of infrastructural planning processes as well
as for the scientific evaluation of new ideas.

Microscopic traffic simulations employ simulation models, which account
for each single vehicle as a separate entity with a defined position
and speed at each moment \cite{treiber2012traffic}. In each simulation
step, the vehicle's individual state is updated using a rule, which
determines the vehicle's acceleration depending on its current environment.
For many car-following models, it is assumed that the chosen acceleration
is a function of the current positions and speeds of the vehicle itself
and of its immediate leader vehicle, i.e., the vehicle driving ahead. 

In this paper we study a popular car-following model, named after
its inventor P.G.~Gipps \cite{gipps_behavioural_1981}, who proposed
the model in 1981. The model is based on the intuitive principle that
drivers drive at the highest speed that ensures not crashing into
the rear of the leader vehicle if it brakes suddenly. This clear approach
for the model derivation gives a comprehensible meaning to each parameter
and is one reason for the model's popularity until today. 

The original model, or one of its variants, is employed in several
important microscopic traffic simulators \cite{url_paramics,aimsun_url,SUMO2012}. 

Although the Gipps model uses an acceleration function, which is derived
from a safety principle {[}see Eqn.~(\ref{eq:Gipps-safety-eqn})
below{]}, it is well-known that for certain choices of parameters
the model exhibits collisions. These collisions cannot be tied to
common causes for collisions in reality but are rather a defect of
the model itself. This poses problems for its application in microsimulators
and different strategies were proposed for the solution \cite{ciuffo_thirty_2012,Krauss1997,Kraus1998}.
All strategies known to the authors constrain the allowed model parameters,
thereby excluding values that seem reasonable to assume under adequate
circumstances given their proposed meaning.

More precisely, if the follower vehicle is disposed to brake significantly
harder than the leader, the model may produce unsafe behavior even
if the follower's estimate of the maximal deceleration of the leader
vehicle is correct. Wilson argued that such a parametrization is unphysical
\cite{wilson_analysis_2001}. Although this may be formally correct
for the original, we argue that a maximal safe following speed should
exist in that situation. We propose that, instead of restricting the
model parameters, one should modify the safety principle used as the
starting point of its derivation. In its original form, the safety
principle only requires the hypothetical gap between the leader and
follower to be larger than zero when both have come to a stop. This
hypothetical gap is derived by assuming braking maneuvers of leader
and follower vehicles during which their hypothetical trajectories
are decoupled. However, if the follower may brake harder than the
leader an intersection of the hypothetical trajectories may occur,
which is disregarded by Gipps' original safety principle and is the
reason for the observed collisions.

In this paper we show that a natural extension of the original Gipps
model exists, which is obtained when Gipps' safety principle is applied
to all time points of the hypothetical braking maneuvers, and which
prevents the described type of collisions. To this end, we briefly
review the original model in Sec.~\ref{sec:Gipps-original-model}
and review its derivation paying a special attention to the problematic
case leading to collisions in Sec.~\ref{sec:Gipps-calculus-reviewed}.
In Section~\ref{sec:vsafe-for-b-ge-bhat} we present the detailed
derivation of an extended model accounting for that case appropriately.
Section~\ref{sec:The-equilibrium-flow} studies the equilibrium flow
yielding an natural extension of the speed-headway function of the
original model. The stability of the equilibrium flow in a many vehicle
system is studied analytically in Sec.~\ref{sec:Stability-of-the-equilibrium}
for identical vehicles and for a heterogeneous ensemble in Section~\ref{sec:Simulation-Experiments}.
In the concluding Section~\ref{sec:Discussion} we provide additional
interpretation for our results and discuss open points.

\section{Gipps' Original Car-Following Model\label{sec:Gipps-original-model}}

The original Gipps model \cite{gipps_behavioural_1981} maps the velocity
$v(t)$ of a driver-vehicle unit to its speed $v(t+\tau)$ attained
at time $t+\tau$ as 
\begin{equation}
v(t+\tau)=v(t)+\tau\cdot a(t).\label{eq:Gipps-update-speed}
\end{equation}
The positional increment is then calculated by a trapezoid rule
\begin{equation}
x(t+\tau)=x(t)+\tau\frac{v(t)+v(t+\tau)}{2},\label{eq:Gipps-update-pos}
\end{equation}
corresponding to a constant value for acceleration $a(s)\equiv a(t)=(v(t+\tau)-v(t))/\tau$
during the interval $s\in[t,t+\tau]$. For the acceleration Gipps
assumes the form
\begin{equation}
a(t)=\min(a_{\max},a_{{\rm safe}})\label{eq:accel-Gipps}
\end{equation}
Here $a_{{\rm max}}=a_{{\rm max}}(v(t))$ is a velocity-dependent
maximal acceleration and $a_{{\rm safe}}$ is a safe acceleration
calculated in dependence of the current speeds and positions of the
vehicle and its immediate leader vehicle, as well as several parameters
described in more detail below. The maximal acceleration as defined
by Gipps is 
\begin{equation}
a_{{\rm max}}(v)=2.5\alpha(1-\frac{v}{v_{{\rm max}}})\sqrt{0.025+\frac{v}{v_{{\rm max}}}},\label{eq:a-max-gipps}
\end{equation}
with a parameter $\alpha>0$ describing the maximal desired acceleration
of the driver. The safe acceleration $a_{{\rm safe}}$ is determined
from a safety equation, which attempts to assure that the vehicle
can break in time if the leading vehicle starts breaking at time $t$
with an assumed deceleration $\hat{B}>0$. Gipps formulates the safety
equation assuming that the ego vehicle accelerates until $t+\tau$
with $a_{{\rm safe}}$, then travels with velocity $v(t+\tau)$ until
$t+\tau+\theta$ and starts breaking with a maximal desired deceleration
$B>0$ afterwards. Here $\theta$ is an additional safety margin often
disregarded or set to a constant proportion with respect to $\tau$.
Gipps himself performed most of his studies only for the case $\theta=\tau/2$.
In this work we keep $\theta$ as a free parameter. Although it does
not have an independent effect on the desired gap $g_{0}$ {[}see
Eq.~(\ref{eq:safe-gap-gipps}){]}, which depends only on the sum
$\tau+\theta$ {[}see Eqn.~(\ref{eq:safe-gap-gipps}) below{]}, the
value of $\theta$ \emph{does} change the dynamic properties of the
model. For instance it has an independent influence on the stability
of the equilibrium flow solution \cite{wilson_analysis_2001}. Moreover,
it affects the approaching and stopping behavior of a follower vehicle.
For instance, a larger value of $\theta$ induced a smooth, or ``creeping'',
stopping, while for $\theta=0$, the stopping is performed to the
point \cite{treiber_comparing_2015}.

The resulting safety equation proposed by Gipps is:

\begin{equation}
\underbrace{\frac{v_{f}(t)+v_{f}(t+\tau)}{2}\cdot\tau}_{\begin{array}{c}
\text{follower's distance}\\
\text{covered in }[t,t+\tau]
\end{array}}+\underbrace{v_{f}(t+\tau)\theta}_{\begin{array}{c}
\text{follower's distance}\\
\text{covered in }[t+\tau,t+\tau+\theta]
\end{array}}+\underbrace{\frac{v_{f}(t+\tau)^{2}}{2B}}_{\begin{array}{c}
\text{follower's}\\
\text{brake-dist}
\end{array}}\le\underbrace{g(t)}_{\begin{array}{c}
\text{current}\\
\text{gap}
\end{array}}+\underbrace{\frac{v_{\ell}(t)^{2}}{2\hat{B}}}_{\begin{array}{c}
\text{leader's}\\
\text{brake-dist}
\end{array}},\label{eq:Gipps-safety-eqn}
\end{equation}
where $v_{f}$ and $v_{\ell}$ are the velocities of the following
and leading vehicles and 
\begin{equation}
g(t)=x_{\ell}^{{\rm back}}-x_{f}^{{\rm front}}-g_{{\rm stop}}\label{eq:def-gap}
\end{equation}
 is their front-to-back gap diminished by a minimal gap $g_{{\rm stop}}>0$,
which should be maintained at standstill. Here $\hat{B}$ denotes
the maximal expected deceleration of the leader as estimated by the
follower.

Gipps proposed the following expression\footnote{Note that Gipps did set $\theta=\tau/2$ in at some point of the calculations
in his original work.} for the maximal safe following speed $v_{{\rm safe}}$ at time $t+\tau$
obtained from equating both sides of (\ref{eq:Gipps-safety-eqn})
and solving for $v(t-\tau)\ge0$:
\begin{equation}
v_{{\rm safe}}=-B(\frac{\tau}{2}+\theta)+\sqrt{\left(B(\frac{\tau}{2}+\theta)\right)^{2}+B\left(2g(t)-v_{f}(t)\tau+\frac{v_{\ell}(t)^{2}}{\hat{B}}\right)}.\label{eq:vsafe-gipps}
\end{equation}
The obtained value $v_{{\rm safe}}$ is then used for the dynamical
update (\ref{eq:Gipps-update-speed})-(\ref{eq:Gipps-update-pos}).
Assuming $v_{{\rm safe}}\ge0$ requires that 
\begin{align}
\underbrace{v_{f}(t)\frac{\tau}{2}}_{\begin{array}{c}
\text{min.dist.}\\
\text{covered till }t+\tau
\end{array}} & \le\underbrace{g(t)}_{\begin{array}{c}
\text{current}\\
\text{gap}
\end{array}}+\underbrace{\frac{v_{\ell}(t)^{2}}{2\hat{B}}}_{\begin{array}{c}
\text{leader's}\\
\text{brake-dist}
\end{array}}.\label{eq:cond-vsafe-ge-0}
\end{align}
An aspect, which is often disregarded but which is quite important
for a well-behaved implementation of the model is the treatment of
the case that (\ref{eq:cond-vsafe-ge-0}) is not fulfilled \cite{treiber_comparing_2015}.
It corresponds to the situation where the follower needs to stop already
within the interval $(t,t+\tau)$ thus braking with $B_{{\rm stop}}$such
that
\[
\underbrace{\frac{v_{f}(t)^{2}}{2B_{{\rm stop}}}}_{\begin{array}{c}
\text{follower's}\\
\text{brake-dist}
\end{array}}=\underbrace{g(t)}_{\begin{array}{c}
\text{current}\\
\text{gap}
\end{array}}+\underbrace{\frac{v_{\ell}(t)^{2}}{2\hat{B}}}_{\begin{array}{c}
\text{leader's}\\
\text{brake-dist}
\end{array}},
\]
i.e., $B_{{\rm stop}}=\frac{b_{\ell}v_{f}(t)^{2}}{2g(t)\hat{B}+v_{\ell}(t)^{2}}.$
This leads to a special case for the update (\ref{eq:Gipps-update-speed})-(\ref{eq:Gipps-update-pos}),
where
\begin{align}
v_{f}(t+\tau) & =0,\label{eq:stop-update-1}\\
x_{f}(t+\tau) & =x_{f}(t)+g(t)+\frac{v_{\ell}(t)^{2}}{2\hat{B}}.\label{eq:stop-update-2}
\end{align}
If stopping within an update step is not implemented (it is not contained
in the original model formulation) the effective deceleration rate
is bounded by $\tilde{B}=\frac{v_{f}\left(t\right)}{\tau}$ since
the vehicle may only decelerate as much as to come to a stop exactly
at time $t+\tau$.

\section{Gipps' Safety Calculus Reviewed\label{sec:Gipps-calculus-reviewed}}

The Gipps model is derived from considerations for the safe following
gap, while seeking to prevent that the follower has to break harder
than using the maximal desired deceleration rate $B$. Gipps already
realized that for a bounded deceleration rate the model does not guarantee
the maintenance of a ``safe speed and distance'' if no further assumptions
are made. Moreover, he gave some conditions when this can be assured
nevertheless. First, it is clear that if the expected maximal deceleration
$\hat{B}$ of the leader is underestimated and the whole calculus
is based on a faulty premise, the model cannot be expected to meet
the desired property. Hence it is required that $\hat{B}\ge B_{\ell}$,
where $B_{\ell}$ corresponds to the actual maximal braking rate applied
by the leader vehicle. Further, Gipps assumed the additional headway
time $\theta$ to equal $\tau/2$, which prevents collisions due to
an inert approach to the desired headway time $\tau+\theta$.

Interestingly, Gipps' definition of ``safe speed and distance''
does not assure the absence of collisions. As it seems, he considers
every speed and distance safe which fulfills Eq.~(\ref{eq:Gipps-safety-eqn})
and his proof only assures that this inequality is invariant under
the dynamics (\ref{eq:Gipps-update-speed})-(\ref{eq:Gipps-update-pos}).
In fact, for the case that the maximal deceleration $B>0$ of the
follower is \emph{larger }than the expected deceleration of a leader
$\hat{B}>0$, the formula (\ref{eq:Gipps-safety-eqn}) does not yield
a guidance for the calculation of a safe distance. Indeed, even for
a constant traveling speed $v_{\ell}\equiv v_{f}\equiv v_{0}$, the
corresponding desired gap 
\begin{equation}
g_{0}=v_{0}\left(\tau+\theta\right)+\frac{v_{0}^{2}}{2}\left(B^{-1}-\hat{B}^{-1}\right),\label{eq:safe-gap-gipps}
\end{equation}
obtained by equating both sides of (\ref{eq:Gipps-safety-eqn}) may
become \emph{negative}. We have $g_{0}<0$ iff
\begin{equation}
B^{-1}-\hat{B}^{-1}<-2\frac{\tau+\theta}{v_{0}}.\label{eq:stat-gap-positive-cond-gipps}
\end{equation}

Anyhow, it seems desirable to account for the case $B>\hat{B}$, e.g.,
to depict traffic situations where drivers choose smaller following
gaps based on their willingness to apply strong braking. In the following
we present strategies to deal with this deficiency of the Gipps model. 

\begin{figure}
\begin{centering}
(a)%
\fbox{\begin{minipage}[t][1\totalheight][b]{0.3\columnwidth}%
\begin{center}
\includegraphics[width=1\textwidth]{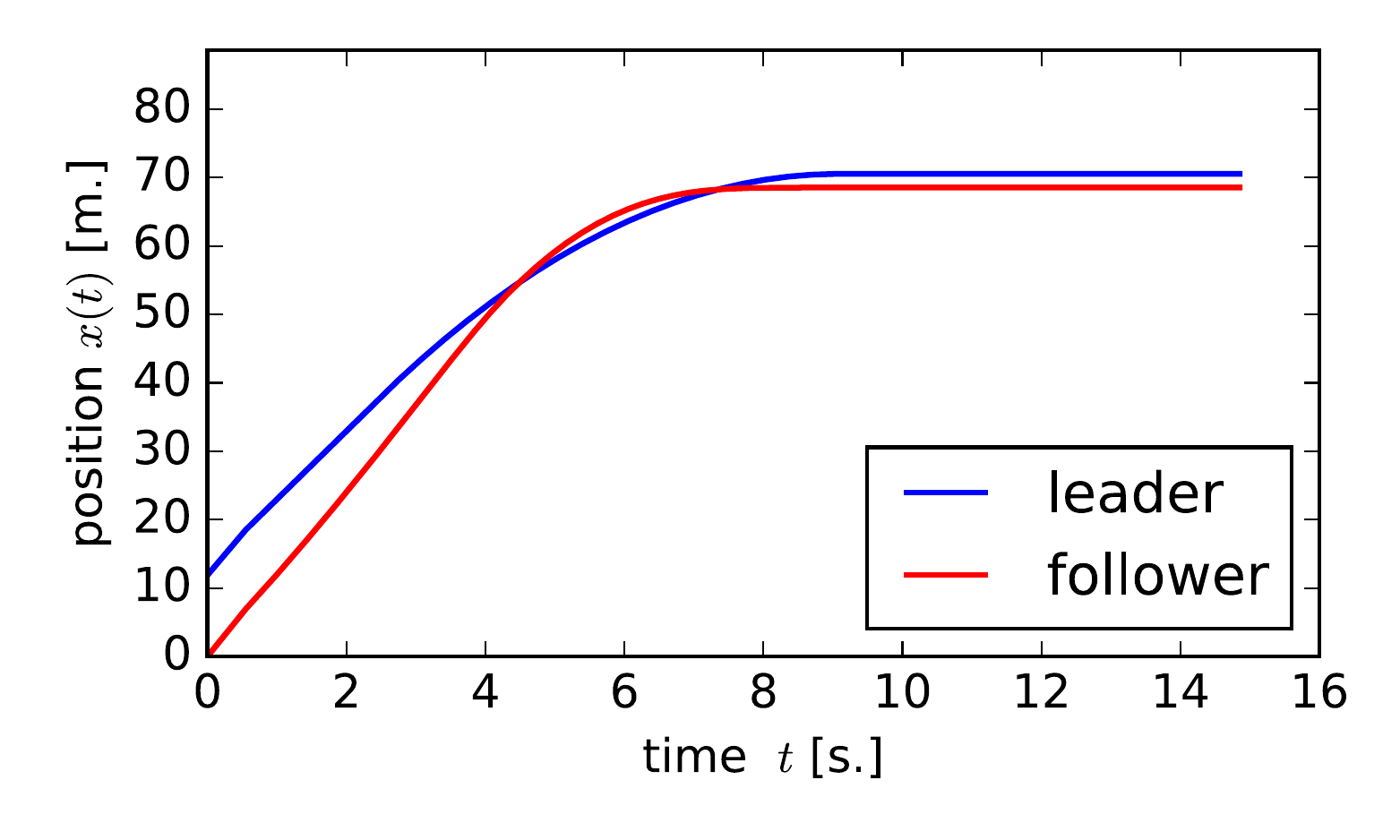}
\par\end{center}
\begin{center}
\includegraphics[width=1\textwidth]{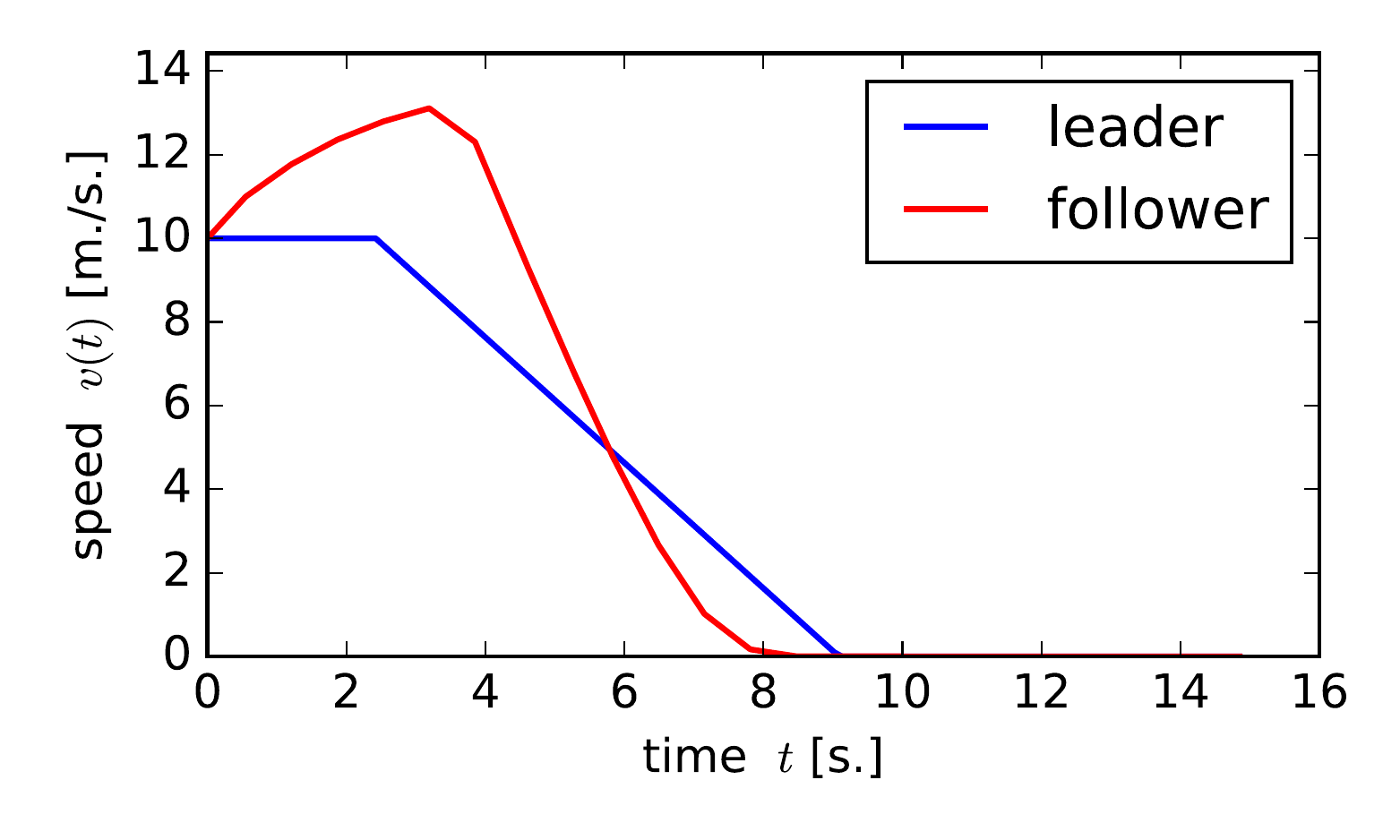}
\par\end{center}%
\end{minipage}}(b)%
\fbox{\begin{minipage}[t][1\totalheight][b]{0.3\columnwidth}%
\begin{center}
\includegraphics[width=1\textwidth]{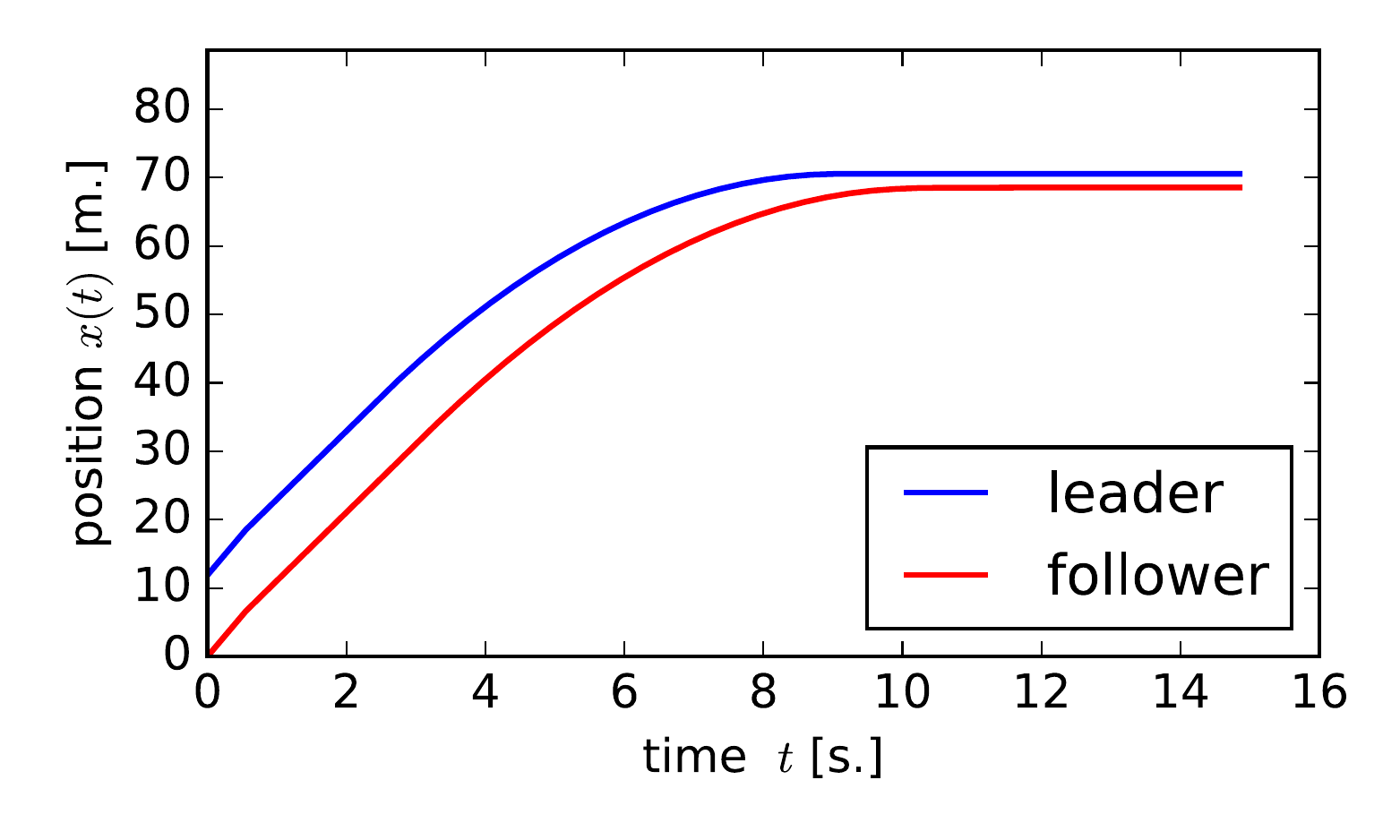}
\par\end{center}
\begin{center}
\includegraphics[width=1\textwidth]{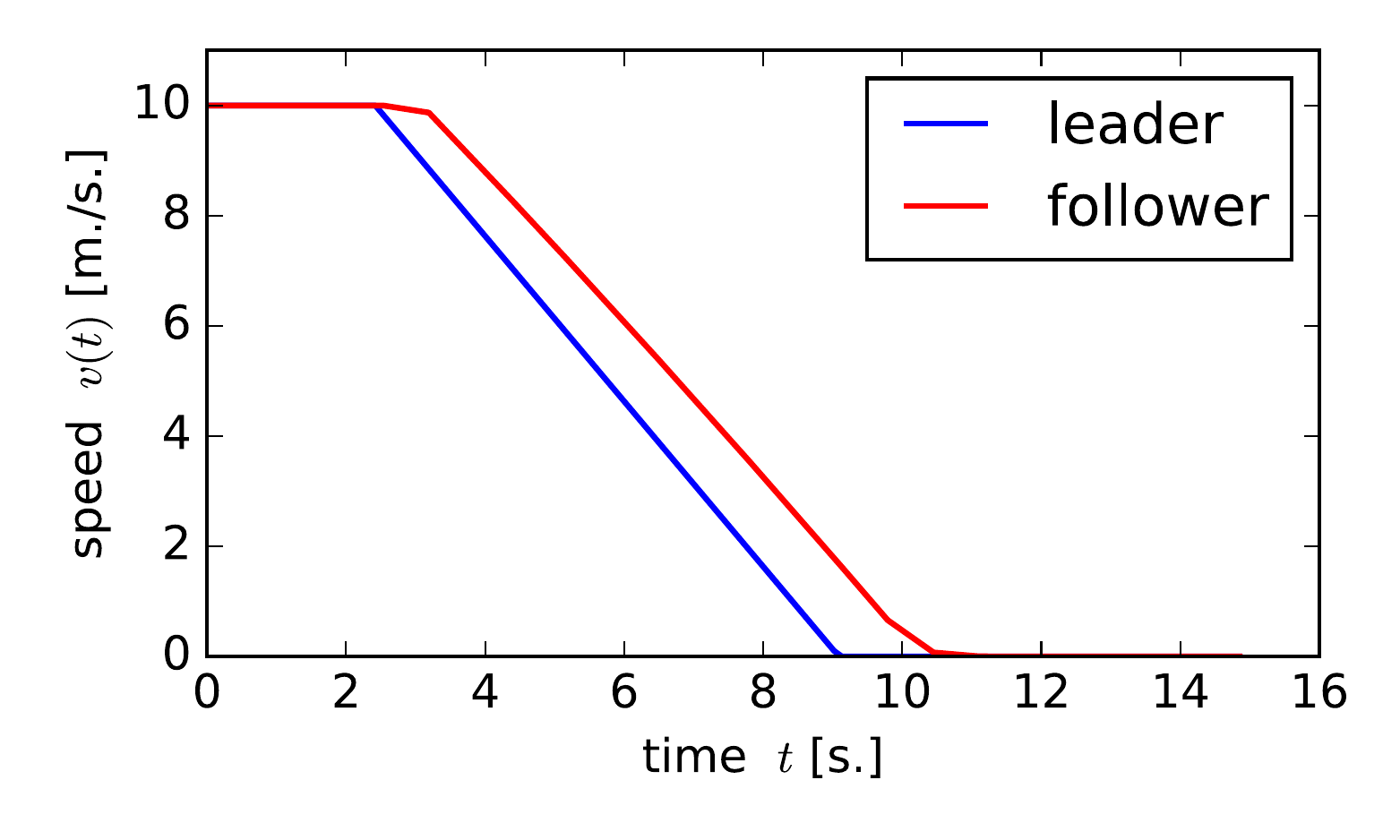}
\par\end{center}%
\end{minipage}}(c)%
\fbox{\begin{minipage}[t][1\totalheight][b]{0.3\columnwidth}%
\begin{center}
\includegraphics[width=1\textwidth]{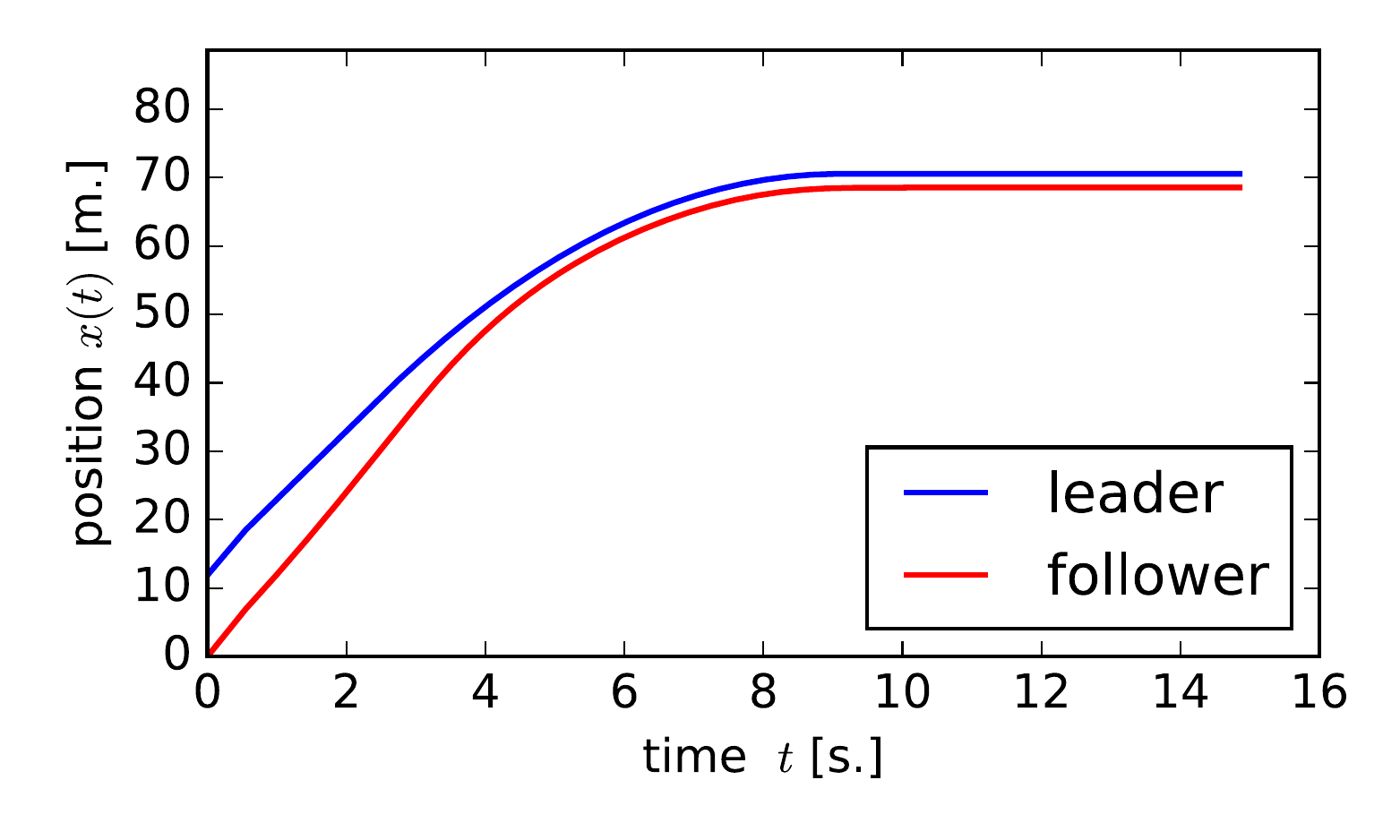}
\par\end{center}
\begin{center}
\includegraphics[width=1\textwidth]{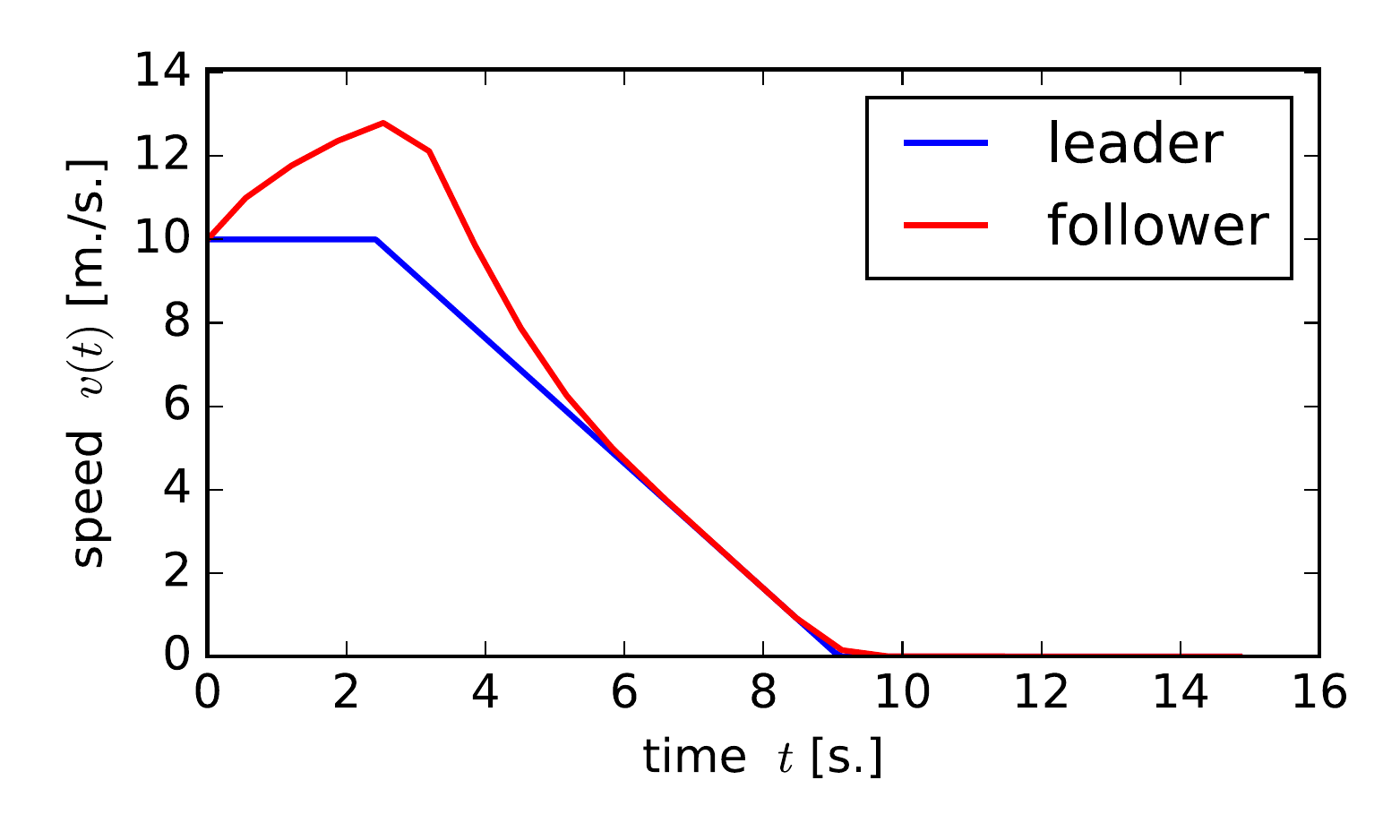}
\par\end{center}%
\end{minipage}}
\par\end{centering}
\caption{\label{fig:Gipps-collision-bf-ge-bl}Comparison of trajectories for
different implementations for the case that the follower has a higher
maximal deceleration rate than the leader, i.e., $B>\hat{B}=B_{\ell}$.
The parameters for all cases are $\tau=0.66{\rm s}$, $\theta=0.33{\rm s}$,
$\hat{B}=B_{\ell}=1.5{\rm m/s^{2}}$, $B=4.5{\rm m/s^{2}}$, and $g_{{\rm stop}}=2{\rm m}$;
The vehicles are inserted at an initial speed of $v_{0}=10{\rm m/s}$
and with an initial temporal headway of $h=\vartheta+g_{{\rm stop}}/v_{0}=1.2{\rm s}$,
which would be stationary for $\hat{B}=B$. The leader vehicle travels
at a speed of $10{\rm m/s}$. At $x\approx50{\rm m}$ it starts breaking
at rate $B_{\ell}$ until its back bumper comes to halt at $x_{{\rm stop}}=80{\rm m}$
(red trajectory). The follower vehicle (blue trajectory) is controlled
by a Gipps model {[}with different strategies for $B>\hat{B}$ and
bounded deceleration rates, see main text{]} and its front comes to
a halt at $x=78{\rm m}=x_{{\rm stop}}-g_{{\rm stop}}$.}
\end{figure}

Figure~\ref{fig:Gipps-collision-bf-ge-bl}~(a) illustrates a collision
for the Gipps model due to $B>\hat{B}$. The leader vehicle first
drives at a constant speed $v_{0}=10{\rm m/s}$ and starts braking
with $\hat{B}=1.5{\rm m/s}$ until it comes to a halt. The follower
vehicle (blue trajectory along its front bumper) is controlled by
a Gipps model with a maximal deceleration bounded to $B=4.5$. It
collides with the rear end of the leader (red trajectory along its
back bumper) at $t\approx6{\rm s}$. After the collision a constant
deceleration rate $B$ is applied. Note that the vehicle comes to
a halt at a position well before the leader's stop position as required
by (\ref{eq:Gipps-safety-eqn}).

In Fig.~\ref{fig:Gipps-collision-bf-ge-bl}~(b) the lack of safety
is resolved by artificially increasing the deceleration rate taken
into account for the leader in (\ref{eq:vsafe-gipps}) to 
\begin{equation}
\hat{B}_{{\rm \max}}=\max\{B,\hat{B}\}.\label{eq:resolution-SUMO}
\end{equation}
This effectively ignores the peculiarities of the situation $B>\hat{B}$
{[}see Section~\ref{sec:Simulation-Experiments}{]}. In (c) the collision
is avoided by using an exact formula for the safe following distance
in the case $B>\hat{B}$, which is presented below. This extended
model is capable to produce smaller, though safe, headways for the
situation $B>\hat{B}$. 

\section{An Extension of the Gipps Model for $B>\hat{B}$\label{sec:vsafe-for-b-ge-bhat}}

In this section we give here an exact formula for the safe speed for
the situation $B>\hat{B}$, i.e. $\Delta B=\hat{B}-B<0$. In this
approach we follow Gipps' original safety concept, which was based
on Eqn.~(\ref{eq:Gipps-safety-eqn}), as close as possible and extend
it by the requirement that $g(t)>0$ can be achieved without braking
harder than $B$. As argued above {[}see Eqs.~(\ref{eq:safe-gap-gipps})-(\ref{eq:stat-gap-positive-cond-gipps}){]}
the latter is not assured by satisfying (\ref{eq:Gipps-safety-eqn})
if $B>\hat{B}$.

\begin{figure}
\centering{}\includegraphics[width=0.8\textwidth]{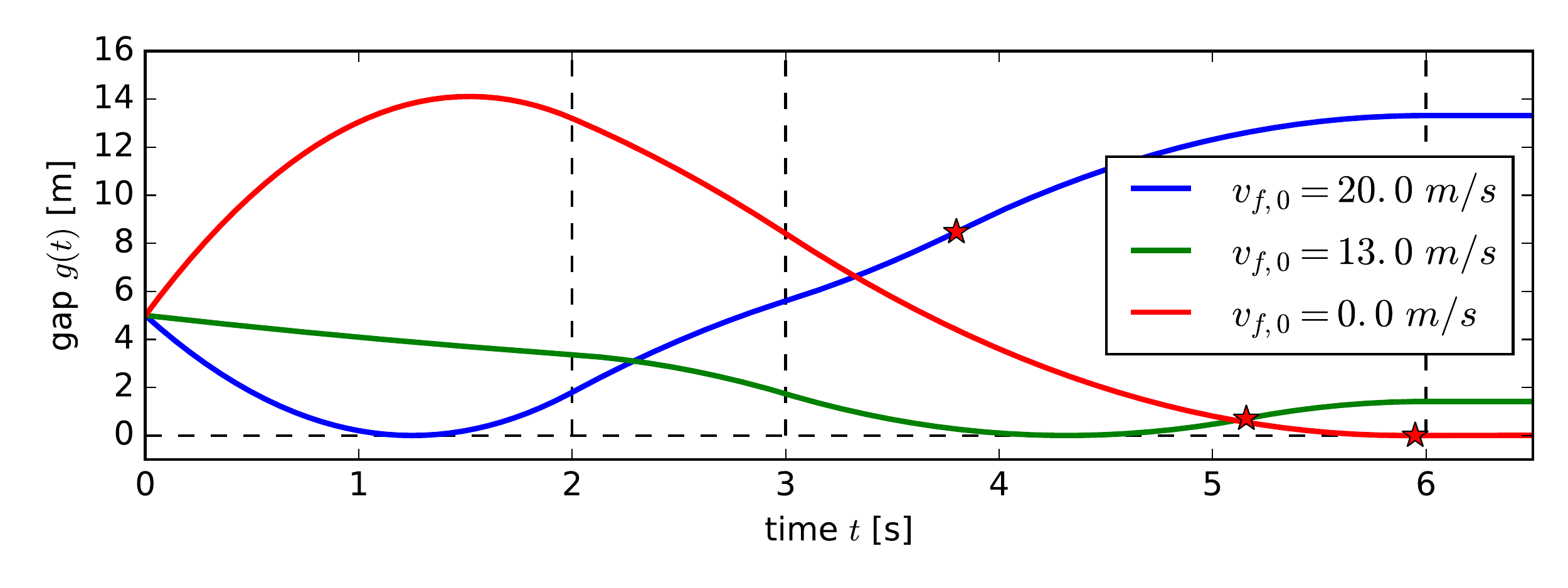}\caption{\label{fig:hypothetical-trajectory-types}
Three different hypothetical
evolutions for the gap $g\left(t\right)$ illustrating the different
possible limiting cases for the follower's acceleration $\alpha$
for $t\in I_{0}=\left(0,\tau\right)$. The different cases are triggered
by varying the initial speed $v_{f,0}$ of the follower, while keeping
constant the initial speed $v_{\ell,0}=12{\rm m/s}$ of the leader
and the initial gap $g_{0}=5{\rm m}$. The blue trajectory corresponds
to the case were the follower initially travels with a much larger
speed $v_{f,0}=20{\rm m/s}$ than the leader. Here, the limiting factor
for the acceleration $\alpha$ during $I_{0}$ is the possible tangency
within $I_{0}$. For a lower initial speed $v_{f,0}=13{\rm m/s}$
the situation changes since for this case a tangency at time $t=t_{\parallel,2}>t_{2}$
is prohibiting a higher acceleration. The red trajectory shows the
hypothetical evolution of $g(t)$ when the follower is stopped initially,
i.e., $v_{f,0}=0$. In this case no tangency can occur until the leader
would have stopped at $t_{s,\ell}=6{\rm s}$, when constantly braking
with rate $\hat{B}=2{\rm m/s^{2}}$. The other model parameters were
set to $B=4{\rm m/s^{2}}$, $\tau=2{\rm s}$, and $\theta=1{\rm s}$.
Red star-shaped markers indicate the extrapolated stopping times $t_{f,s}$
of the follower for the different trajectories.}
\end{figure}

To derive the corresponding value of $v_{f}(t+\tau)=v_{{\rm safe}}$
we follow the reasoning sketched in Sec.~\ref{sec:Gipps-original-model},
considering a follower and a leader vehicle described by their positions
and speeds $\left(x_{f},v_{f}\right)$ and $\left(x_{\ell},v_{\ell}\right)$.
Without loss of generality we consider the situation to be initialized
at $t_{0}=0$. The initial state is denoted by 
\[
\left(x_{f,0},v_{f,0}\right):=\left(x_{f}\left(0\right),v_{f}\left(0\right)\right),\ \left(x_{\ell,0},v_{\ell,0}\right):=\left(x_{\ell}\left(0\right),v_{\ell}\left(0\right)\right).
\]
As for the original model, the target speed is selected based on the
hypothesized scenario that the leader starts braking with rate $\hat{B}$
at $t=0$. To denote the expected states in this scenario we use notations
$v_{f,1}=v_{f}(t_{1}),$ $x_{\ell,2}=x_{\ell}(t_{2}),$ etc., with
time points 
\[
t_{1}=\tau,\text{ and }t_{2}=\tau+\theta.
\]
As a shorthand, we introduce the following notation for the corresponding
intervals $I_{j}$ of potentially different values for the follower's
acceleration rate $a_{f}$:
\begin{itemize}
\item $I_{0}=\left(0,t_{1}\right)$, where $a_{f}(t)\equiv\alpha$ with
$\alpha$ to be determined,
\item $I_{1}=\left(t_{1},t_{2}\right)$, where $a_{f}(t)\equiv0$,
\item $I_{2}=\left(t_{2},\infty\right)$, where $a_{f}(t)\equiv-B$. 
\end{itemize}
The computations in this section are more involved than for the case
$B\le\hat{B}$ because more cases have to be distinguished. There
are two types of important events in a collision free situation. Firstly,
the expected stopping times $t_{s,f}$ and $t_{s,\ell}$ of the vehicles.
Here, the leader's stopping time (when constant braking with rate
$\hat{B}$ is assumed) is 
\begin{equation}
t_{s,\ell}=\frac{v_{\ell,0}}{\hat{B}}.\label{eq:t-sl}
\end{equation}
In addition to the stopping times the case $B>\hat{B}$ allows for
another type of event, namely a tangency of the vehicles' trajectories
while both are moving at positive speed, i.e., at some moment $t=t_{\parallel}<\min\left\{ t_{s,f},t_{s,\ell}\right\} $.
The occurrence of such a tangency can be determined from the condition
\begin{equation}
g(t_{\parallel})=g^{\prime}(t_{\parallel})=0.\label{eq:tangency-def}
\end{equation}
Note that the reason for the absence of tangencies in the case $B\le\hat{B}$
is that the evolution of the gap $g(t)$ is convex, i.e. $g^{\prime\prime}(t)\le0$,
for $t<t_{s,\ell}$ {[}see Fig.~\ref{fig:hypothetical-trajectory-types}{]}.
For $B>\hat{B}$, tangencies may occur within $I_{0}$ and $I_{2}$,
and this implies that more than one type of upper bound on the value
of the acceleration $\alpha$ within $I_{0}$ has to be considered.
Figure~\ref{fig:hypothetical-trajectory-types} shows the three different
cases that may occur (excluding cases, where a stop within $I_{0}$
is necessitated).

A necessary condition for the existence of a tangency at time $t_{\parallel,j}$
in an interval $I_{j}$, $j=0,2$, is that the gap is concave in $I_{j}$
and decreases at interval begin. Therefore no tangency occurs in $I_{1}$,
where the gap is always convex. A potential tangency at time $t_{\parallel,j}<t_{s,\ell}$
defines an upper bound $\alpha_{j}$ for $\alpha$. Formally, we can
even obtain $t_{\parallel,j}>t_{s,\ell}$ if the vehicles' deceleration
is extrapolated beyond their stopping times, such that the tangency
would occur for negative values of the speed. In that case, the reasoning
according to (\ref{eq:Gipps-safety-eqn}) has to be applied to obtain
the appropriate bound $\alpha_{j}$ on $\alpha$, i.e., $\alpha_{j}=\left(v_{{\rm safe}}-v_{f,0}\right)/\tau$
with $v_{{\rm safe}}$ from (\ref{eq:vsafe-gipps}), or $\alpha_{j}=B_{{\rm stop}}$
corresponding to the stopping update (\ref{eq:stop-update-1})\textendash (\ref{eq:stop-update-2}).
In the following we will derive the constraints for the intervals
$I_{0}$ and $I_{2}$, i.e., $\alpha_{0}$ and $\alpha_{2}$.

\textbf{Interval $I_{0}$}. The expected evolution of the gap $g(t)$
for $t\le\tau$ is
\begin{equation}
g(t)=g_{0}+g_{0}^{\prime}t-\left(\alpha+\hat{B}\right)\frac{t^{2}}{2},\label{eq:gap-I0}
\end{equation}
with $g_{0}^{\prime}=g^{\prime}(0)=v_{\ell,0}-v_{f,0}$ and $g_{0}=g(0)\ge0$.
A tangency for the vehicle's trajectories within $I_{0}$ requires
that there exists $t_{\parallel,0}\in I_{0}$ such that $g(t_{\parallel,0})=0$
and 
\[
0=g^{\prime}\left(t_{\parallel,0}\right)=g_{0}^{\prime}-\left(\alpha+\hat{B}\right)t_{\parallel,0}.
\]
Hence, $t_{\parallel,0}=\frac{g_{0}^{\prime}}{\alpha+\hat{B}}.$ Inserting
this into (\ref{eq:gap-I0}) and solving $g(t_{\parallel,0})=0$ for
$\alpha$ yields
\begin{equation}
\alpha_{0}=-\frac{g_{0}^{\prime2}}{2g_{0}}-\hat{B}.\label{eq:alpha-I0}
\end{equation}
Thus,
\begin{equation}
t_{\parallel,0}=-\frac{2g_{0}}{g_{0}^{\prime}}.\label{eq:tpar-in-I0}
\end{equation}
If $0\le t_{\parallel,0}\le\min\left\{ t_{s,\ell},\tau\right\} $,
it is required that $\alpha\le\alpha_{0}$. If instead $t_{s,\ell}\le\tau$
and $t_{\parallel,0}\notin\left(0,t_{s,\ell}\right)$, we compute
the follower's next step as for the original model. Note that $t_{\parallel,0}>0$
requires $g_{0}^{\prime}<0$. Further, $v_{f,0}+\alpha\tau$ indicates
that a stop is required within $I_{0}$ for the follower. In this
case the following update is to be applied:
\begin{align}
v\left(t+\tau\right) & =0,\label{eq:stopping-in-I0-1}\\
x\left(t+\tau\right) & =x_{0,f}+\frac{v_{f,0}^{2}}{2\alpha_{0}}.\label{eq:stopping-in-I0-2}
\end{align}

\textbf{Interval $I_{1}$}. First, we observe that if the leader has
come to a halt at the end of $I_{1}$, i.e., $t_{s,\ell}<t_{2}$,
further tangencies can be disregarded as they could only occur in
$I_{2}$ and the classical Gipps reasoning may be applied to obtain
a bound $\alpha_{2}=\left(v_{{\rm safe}}-v_{f,0}\right)/\tau$ with
$v_{{\rm safe}}$ according to (\ref{eq:vsafe-gipps}). Note that
the constraint $\alpha\le\alpha_{0}$, which is applied if $0\le t_{\parallel,0}\le\min\left\{ t_{s,\ell},\tau\right\} $,
may still induce a different behavior than for the original model
if $\alpha_{0}<\alpha_{2}$.

If no stop must be performed by the follower within $I_{0}$ and $t_{s,\ell}>t_{2}$,
the evolution of the gap $g(t)$ for $t>t_{1}=\tau$ has to be considered
in more detail. For $t\in I_{1}$ the follower is assumed to proceed
with constant speed 
\[
v_{f,1}=v_{f,0}+\tau\alpha,
\]
while the leader keeps on braking with rate $\hat{B}$. Hence,

\begin{equation}
g(t_{1}+s)=g_{1}+g_{1}^{\prime}s-\hat{B}\frac{s^{2}}{2},\label{eq:gap-I1}
\end{equation}
where
\begin{align}
g_{1}^{\prime}=g^{\prime}(t_{1}) & =g_{0}^{\prime}-\left(\hat{B}+\alpha\right)\tau,\label{eq:gp1}\\
g_{1}=g(t_{1}) & =g_{0}+\left(g_{0}^{\prime}+g_{1}^{\prime}\right)\frac{\tau}{2}.\label{eq:g1}
\end{align}
At the end of $I_{1}$, the gap's state has evolved to 
\begin{align}
g_{2}^{\prime}=g^{\prime}(t_{2}) & =g_{1}^{\prime}-\hat{B}\theta\label{eq:gp2}\\
g_{2}=g(t_{2}) & =g_{1}+\left(g_{1}^{\prime}+g_{2}^{\prime}\right)\frac{\theta}{2}\label{eq:g2}
\end{align}

\textbf{Interval $I_{2}$}. Within $I_{2}$ it is assumed that both
vehicles decelerate with the rates $B$ and $\hat{B}$, respectively.
Thus, 
\begin{equation}
g(t_{2}+s)=g_{2}+g_{2}^{\prime}s+\left(B-\hat{B}\right)\frac{s^{2}}{2}.\label{eq:gap-I2}
\end{equation}
Since $B>\hat{B}$, the gap $g(t)$ is concave for $t>t_{2}$ until
the follower stops. Consequently, a tangency may occur for an appropriate
choice of $\alpha$. The requirement $g^{\prime}\left(t_{\parallel,2}\right)=0$
yields 
\begin{equation}
t_{\parallel,2}=t_{2}+\frac{g_{2}^{\prime}}{\hat{B}-B}.\label{eq:tpar2}
\end{equation}
Note that $t_{\parallel,2}\ge t_{2}$ requires that 
\begin{equation}
g_{2}^{\prime}\le0.\label{eq:gp2-le-0}
\end{equation}
Using (\ref{eq:tpar2}) in $g\left(t_{\parallel,2}\right)=0$ gives
\[
g_{2}=\frac{g_{2}^{\prime2}}{2\left(B-\hat{B}\right)}.
\]
Substituting (\ref{eq:gp2})-(\ref{eq:g2}) and applying the quadratic
formula for $g_{1}^{\prime}$ gives
\begin{equation}
g_{1,\ast}^{\prime}:=\frac{1}{2}\left(B-\hat{B}\right)\tau+B\theta-\frac{1}{2}\,\sqrt{\left(B-\hat{B}\right)^{2}\tau^{2}+4\left(B-\hat{B}\right)\left(\left(\tau\theta+\theta^{2}\right)B+\mathit{g}_{0}^{\prime}\tau+2g_{0}\right)}.\label{eq:slns-gp1}
\end{equation}
Here the lower branch of solutions was selected since $g_{1}^{\prime}\le\hat{B}\theta$
is required from (\ref{eq:gp2-le-0}) and (\ref{eq:gp2}). Equation~(\ref{eq:slns-gp1})
yields a valid solution $g_{1,\ast}^{\prime}\le\hat{B}\theta$ as
long as 
\begin{equation}
0\le\hat{B}\left(\theta\tau+\theta^{2}\right)+\mathit{g}_{0}^{\prime}\tau+2g_{0}\label{eq:cond-existence-tangency-in-I2}
\end{equation}
From (\ref{eq:tpar2}) we obtain
\begin{equation}
t_{\parallel,2}=t_{2}+\frac{g_{1,\ast}^{\prime}-\hat{B}\theta}{\hat{B}-B}\label{eq:tpar-in-I2}
\end{equation}
and if $t_{\parallel,2}<t_{s,\ell}$ and Eq.~(\ref{eq:cond-existence-tangency-in-I2})
holds, the potential tangency in $I_{2}$ acts limiting on $\alpha$
implying an upper bound
\begin{equation}
\alpha_{2}=\frac{g_{0}^{\prime}-g_{1,\ast}^{\prime}}{\tau}-\hat{B},\label{eq:alpha2}
\end{equation}
which is obtained by inserting $g_{1,\ast}^{\prime}$ into (\ref{eq:gp1}).

\textbf{Algorithmic choice of $v_{{\rm safe}}$}.\textbf{ }Summarizing
the calculations of this section, we have developed the following
procedure to calculate $v_{{\rm safe}}$ for the case $B>\hat{B}$:
\begin{enumerate}
\item Define $v_{f,0}=v_{f}(t)$, $v_{\ell,0}=v_{\ell}\left(t\right),$
and $g_{0}=x_{\ell}(t)-x_{f}(t)$.
\item Initialize a variable $\alpha\leftarrow\infty$ {[}Eqn.~(\ref{eq:a-max-gipps}){]}.
\item Determine $t_{s,\ell}$ {[}Eqn.~(\ref{eq:t-sl}){]}
\item Determine $t_{\parallel,0}$ {[}Eqn.~(\ref{eq:tpar-in-I0}){]}
\begin{enumerate}
\item If $t_{\parallel,0}\in\left(0,t_{s,\ell}\right),$ set $\alpha\leftarrow\min\left\{ \alpha,\alpha_{\parallel,0}\right\} $
{[}Eqn.~(\ref{eq:alpha-I0}){]} 
\item Otherwise, set $\alpha\leftarrow\min\left\{ \alpha,\left(v_{{\rm safe}}-v_{f,0}\right)/\tau\right\} $
and proceed with step~\ref{enu:calc-vnext}. 
\end{enumerate}
\item Check if a stop is required within $I_{0}$, i.e., $v_{f,0}+\alpha\tau\le0$.
If this is the case we can determine the vehicle's next state immediately:
\begin{enumerate}
\item If $t_{\parallel,0}\in\left(0,t_{s,\ell}\right),$ perform the state
update according to Eqs.~(\ref{eq:stopping-in-I0-1})\textendash (\ref{eq:stopping-in-I0-2})
\item Otherwise, perform the state update according to Eqs.~(\ref{eq:stop-update-1})\textendash (\ref{eq:stop-update-2})
\end{enumerate}
\item Determine $t_{\parallel,2}$ {[}Eqn.~(\ref{eq:tpar-in-I2}){]}
\begin{enumerate}
\item If $t_{\parallel,2}\in\left(t_{2},t_{s,\ell}\right),$ set $\alpha\leftarrow\min\left\{ \alpha,\alpha_{\parallel,2}\right\} $
{[}Eqn.~(\ref{eq:alpha2}){]}
\item Otherwise, set $\alpha\leftarrow\min\left\{ \alpha,\left(v_{{\rm safe}}-v_{f,0}\right)/\tau\right\} $
\end{enumerate}
\item Determine the safe and maximal speed {[}Eqn.~(\ref{eq:a-max-gipps}){]}
as 
\begin{align}
v_{{\rm safe}} & =v_{f,0}+\tau\alpha.\label{eq:safe-next-speed-extended}\\
v_{{\rm max}} & =v_{f,0}+\tau a_{{\rm max}}(v_{f,0})\label{eq:max-next-speed}
\end{align}
\item \label{enu:calc-vnext}Use (\ref{eq:safe-next-speed-extended})\textendash (\ref{eq:max-next-speed})
to update the speed 
\begin{equation}
v\left(t+\tau\right)=\min\left\{ v_{{\rm safe}},v_{{\rm max}}\right\} .\label{eq:speed-update-extended}
\end{equation}
\end{enumerate}
We summarize that there are five possible regimes governing a specific
step $\left(x_{f}(t),v_{f}(t)\right)\to\left(x_{f}(t+\tau),v_{f}(t+\tau)\right)$
for the follower:
\begin{enumerate}
\item \label{enu:select-orig-vsafe}The original Gipps $v_{{\rm safe}}$
{[}Eqn.~(\ref{eq:vsafe-gipps}){]} is selected: $\alpha=(v_{{\rm safe}}-v_{f}(t))/\tau$.
\item The vehicle needs to perform a stop within $I_{0}$ according to (\ref{eq:stopping-in-I0-1})\textendash (\ref{eq:stopping-in-I0-2}).
\item \label{enu:select-amax}The maximal acceleration $a_{{\rm max}}(v_{n})$
acts limiting: $\alpha=a_{{\rm max}}(v_{n})$
\item \label{enu:select-alpha0}The possible tangency in $I_{0}$ acts limiting:
$\alpha=\alpha_{\parallel,0}$
\item \label{enu:select-alpha2}The possible tangency in $I_{2}$ acts limiting:
$\alpha=\alpha_{\parallel,2}$
\end{enumerate}
The cases \ref{enu:select-orig-vsafe}\textendash \ref{enu:select-amax}
are already present for the original Gipps model, while \ref{enu:select-alpha0}
and \ref{enu:select-alpha2} are due to the additional class of constraints
imposed by the possibility of tangent trajectories.

\begin{figure}
\begin{centering}
(a)%
\fbox{\begin{minipage}[t][1\totalheight][b]{0.3\columnwidth}%
\begin{center}
\includegraphics[width=1\textwidth]{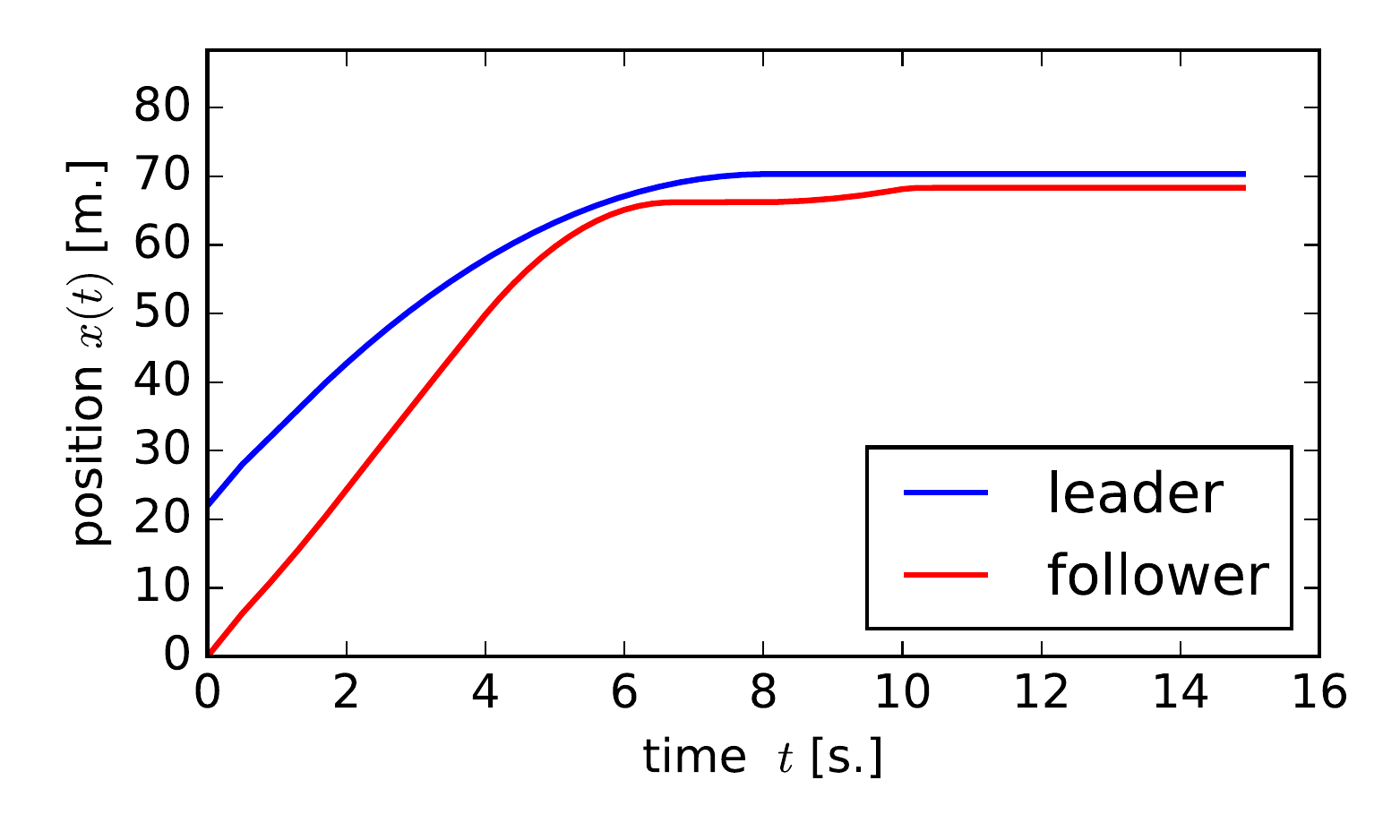}
\par\end{center}
\begin{center}
\includegraphics[width=1\textwidth]{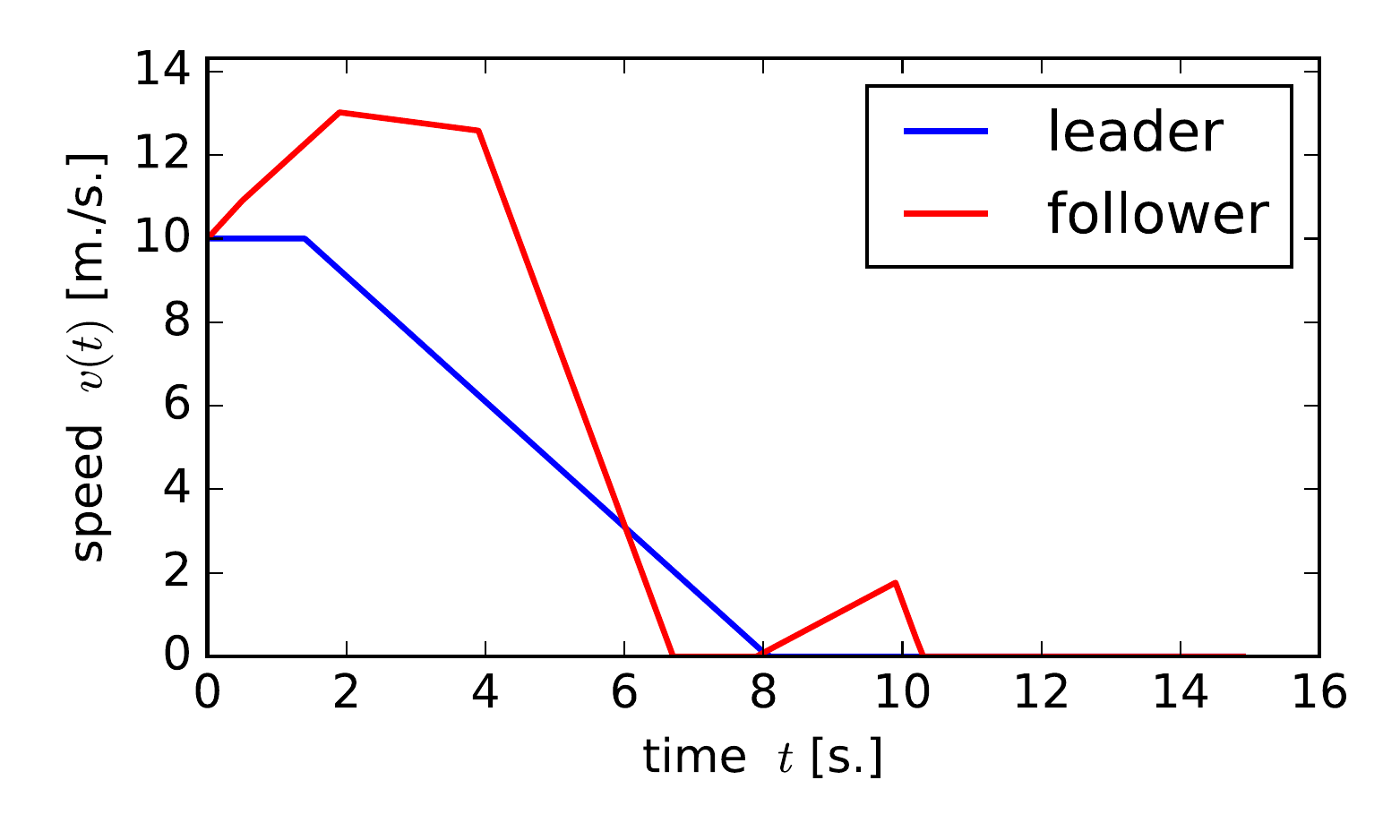}
\par\end{center}%
\end{minipage}}(b)%
\fbox{\begin{minipage}[t][1\totalheight][b]{0.3\columnwidth}%
\begin{center}
\includegraphics[width=1\textwidth]{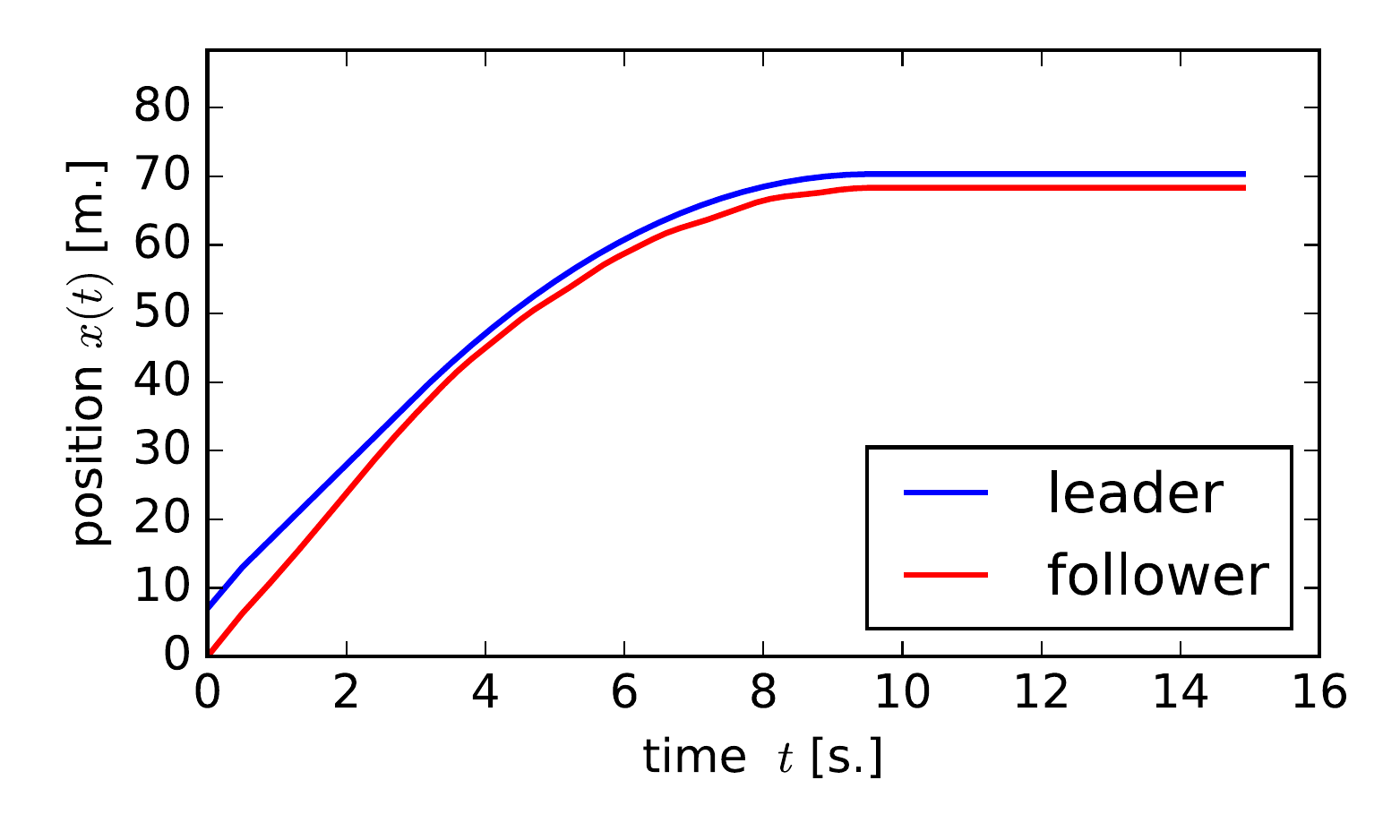}
\par\end{center}
\begin{center}
\includegraphics[width=1\textwidth]{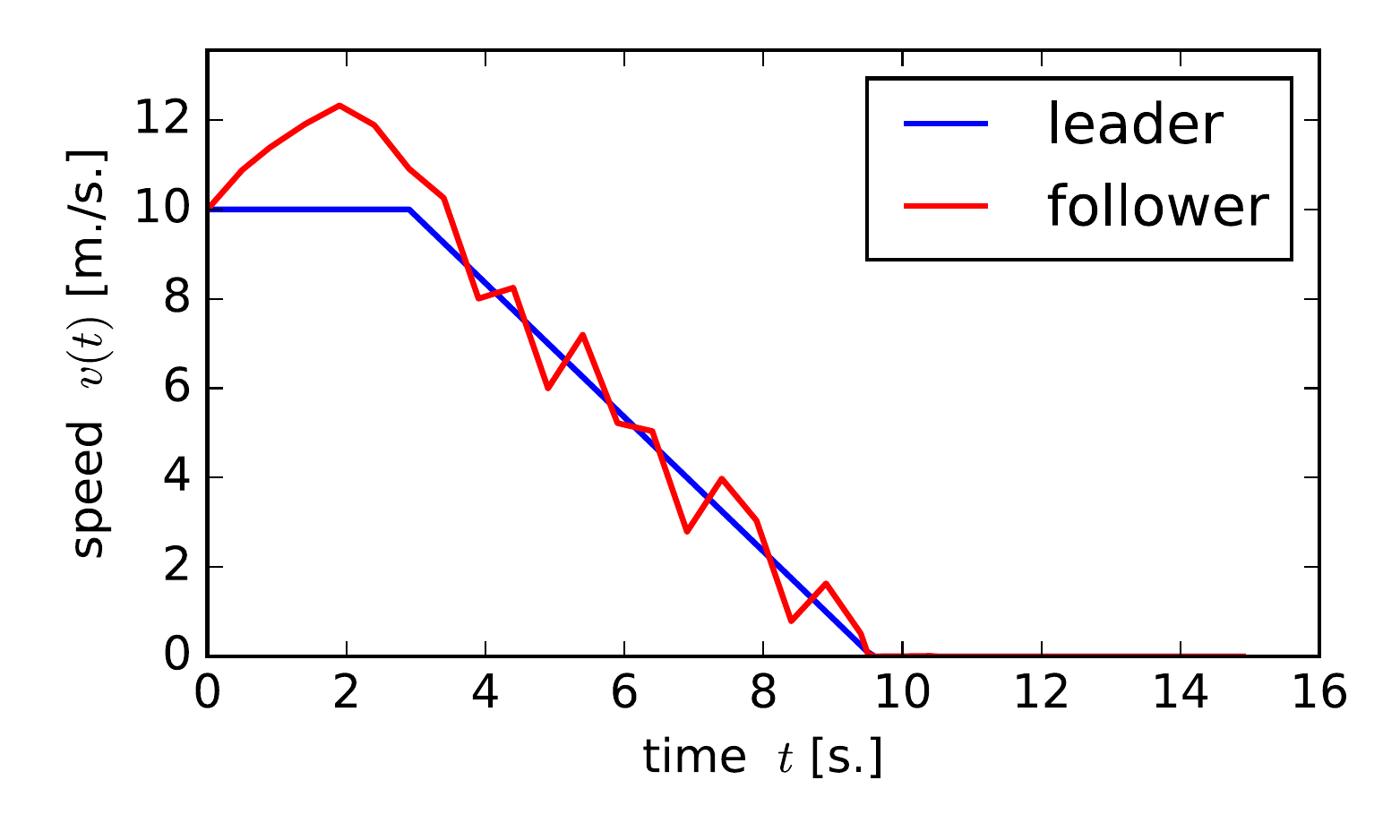}
\par\end{center}%
\end{minipage}}(c)%
\fbox{\begin{minipage}[t][1\totalheight][b]{0.3\columnwidth}%
\begin{center}
\includegraphics[width=1\textwidth]{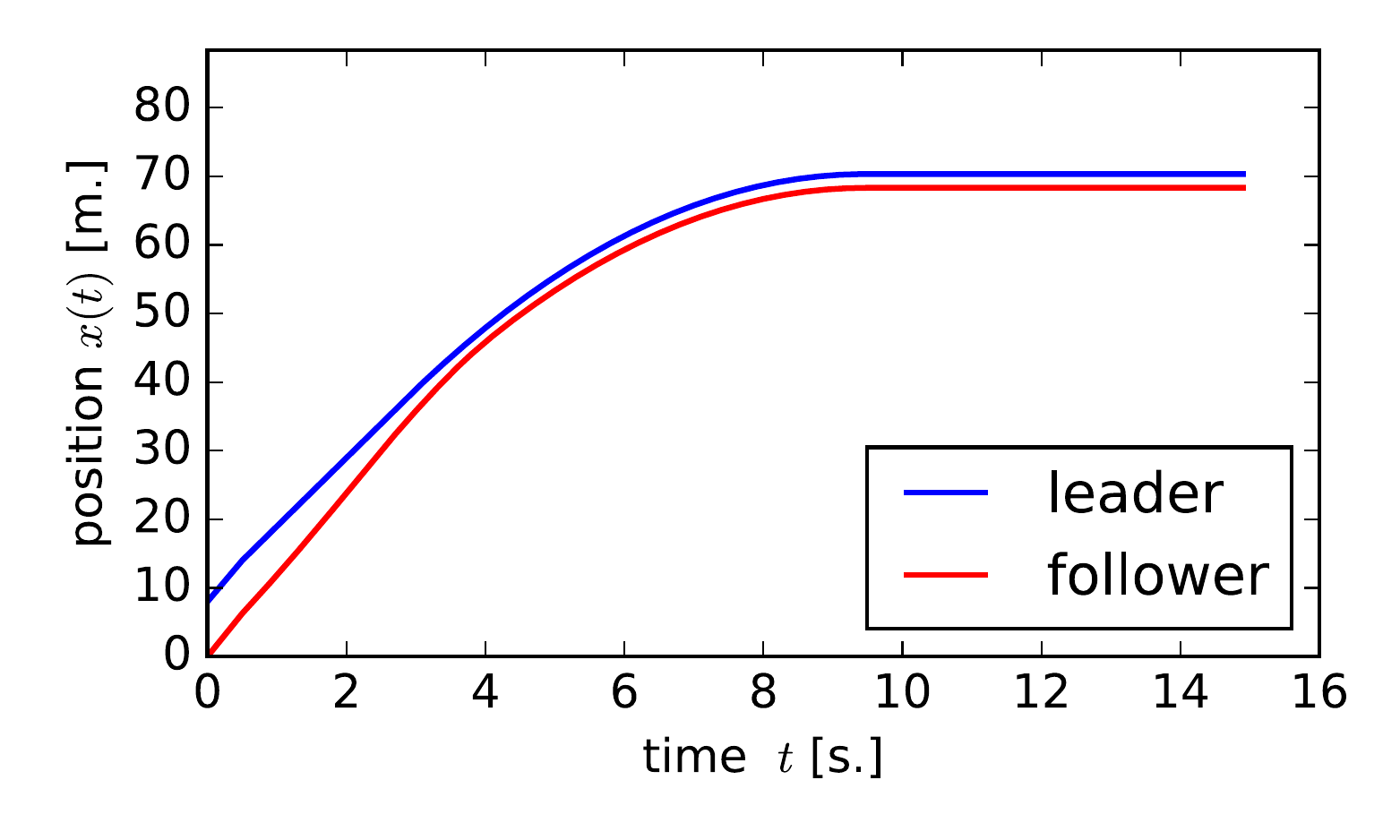}
\par\end{center}
\begin{center}
\includegraphics[width=1\textwidth]{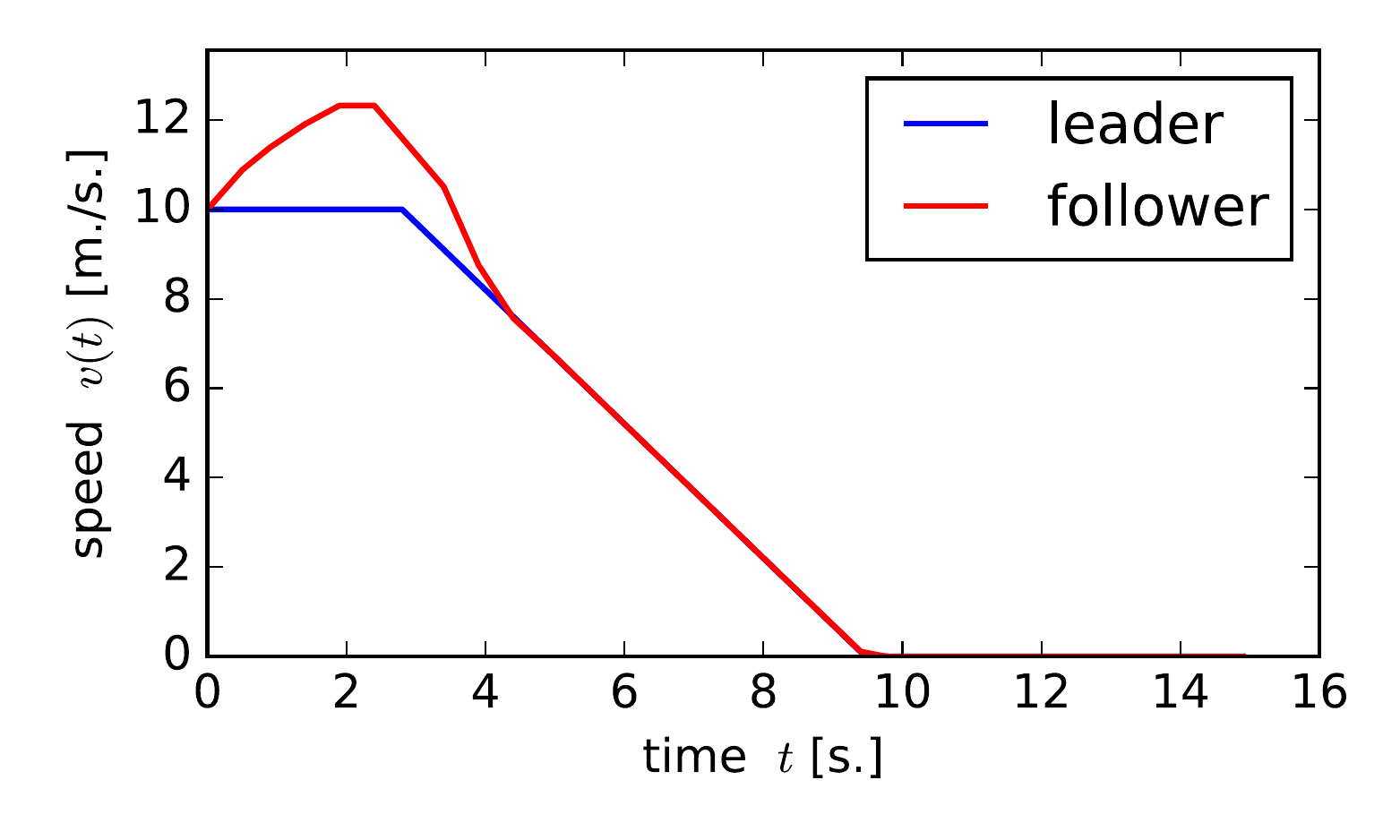}
\par\end{center}%
\end{minipage}}
\par\end{centering}
\begin{centering}
\caption{\label{fig:bouncing}'Bouncing' approach due to repeated tangental
encounters of the follower's and leader's trajectories induced by
too low values for $\theta$ for the proposed extension of the Gipps
model. Panel~(a) shows a large tagental overshoot for $\tau=2$ and
$\theta=0$. Panel~(b) shows that the phenomenon is present for smaller
values of $\tau$ as well, as can be seen from the evolution of the
follower's speed $v_{f}(t)$ (red line in the lower plot). Panel~(c)
illustrates that the 'bouncing' is supressed for non-vanishing $\theta$
(here $\theta=0.1$, $\tau=0.5$). The experimental setup and the
parameters were chosen as in Fig.~\ref{fig:Gipps-collision-bf-ge-bl}
if not stated otherwise. }
\par\end{centering}
\end{figure}

Figure~\ref{fig:Gipps-collision-bf-ge-bl}(c) shows a scenario illustrating
the dynamics of the extended Gipps model, where the follower employs
the tangency avoidance scheme to prevent the collision shown in Fig.~\ref{fig:Gipps-collision-bf-ge-bl}(a).
The example shows a presumably desirable behavior as the follower
smoothly approaches the minimal safe gap. However, it is important
to point out that the parametrization of the model has to be taken
out with some care. In particular, it is recommended to choose the
(non-negative) parameter $\theta$ of sufficient magnitude as very
small values may lead to a 'bouncing' approach to a braking leader
as illustrated in Fig.~\ref{fig:bouncing}(a) and (b). In such a
situation one or several subsequent tangencies can be observed due
to an 'overshoot' leading to a larger gap than the minimal safe following
gap at the end of the update step of length $\tau$. In Panel~(a)
a large overshoot can be observed due to a relatively large value
of $\tau=2$ and vanishing $\theta=0$. Intermediately, it even leads
to a complete stopping. For smaller $\tau$ the effect is still present
though less obvious on the spatial level {[}see $x(t)$ in the upper
plot{]}. Nevertheless, an inspection of the evolution of the speed
{[}$v(t)$ in the lower plot{]} for $\tau=0.5$ shows abrupt changes
of deceleration and acceleration in consecutive steps {[}see Fig.~\ref{fig:bouncing}(b){]}.
The 'bouncing' effect can be effectively ameliorated by introducing
$\theta>0$ {[}see Fig.~\ref{fig:bouncing}(c){]}.

\section{The Equilibrium Flow for the Case $B>\hat{B}$\label{sec:The-equilibrium-flow}}

In this section, we study the equilibrium flow in a system of several
identical vehicle driver units, which obey the dynamics introduced
in Sec.~\ref{sec:vsafe-for-b-ge-bhat}. Note that this implies that
all drivers are disposed to brake with the same maximal comfortable
rate $B$, while they estimate the leader's maximal braking rate by
$\hat{B}<B$. This means that all drivers are systematically underestimating
their leader's disposition for hard braking. Note that this does not
automatically lead to an instability as shown in \cite{wilson_analysis_2001}.

The equilibrium flow is a state of uniform speeds and inter-vehicle
gaps, i.e.,
\[
v_{n}(t)\equiv v_{\ast}\text{, and }g_{n}\left(t\right)\equiv g_{\ast},
\]
where $v_{n}$ is the speed of the $n$-th vehicle and $g_{n}$ is
its front gap. The front gap is defined as the remainder when subtracting
a desired stopping distance $g_{{\rm stop}}$ from the bumper-to-bumper
gap of the $n$-th vehicle and its leader, the $\left(n+1\right)$-th
vehicle. That is
\[
g_{n}=x_{n-1}-l-g_{{\rm stop}}-x_{n},
\]
where $x_{n}$ is the $n$-th vehicle's absolute position and $l$
is the (uniform) vehicle length. A thorough analysis for the original
Gipps model has been taken out by Wilson already \cite{wilson_analysis_2001}.
Therefore we restrict our studies to situations, where the dynamics
governing the equilibrium differ from (\ref{eq:cond-vsafe-ge-0}).
That is, we consider equilibrium solutions, where a possible tangency
limits the safe next speed for the vehicle, see Sec~\ref{sec:vsafe-for-b-ge-bhat}.
For the considerations below, studying the dynamics of the $n$-th
vehicle as a follower of the $\left(n-1\right)$-th vehicle at time
$t$, such that the input parameters for determining the new speed
$v_{n}(t+\tau)$ of the follower are $g_{n}(t)$, $v_{n}(t)$, and
$v_{n-1}(t)$. These are to be identified with $g_{0}$, $v_{f,0}$,
and $v_{\ell,0}$ from Sec.~\ref{sec:vsafe-for-b-ge-bhat}.

At first glance one may think there exist two different cases: one,
where a tangency in $I_{0}$ limits the acceleration and another,
where a tangency in $I_{2}$ poses the decisive constraint. However,
in equilibrium flow we have $g_{n}^{\prime}\left(t\right)\equiv0$.
Hence, no tangency can occur in $I_{0}$ if not $g_{n}\left(t\right)\equiv0$.
We disregard this case in the following and restrict our analysis
to equilibrium regimes $\left(g_{\ast},v_{\ast}\right)$ located within
a neighborhood, where
\begin{equation}
v_{n}\left(t+\tau\right)=v_{n}(t)+\alpha_{\parallel,2}\left(g_{n}\left(t\right),v_{n}\left(t\right),v_{n-1}\left(t\right)\right)\tau,\label{eq:vn-t1}
\end{equation}
for all $v_{n}\left(t\right)$ and $v_{n-1}\left(t\right)$ sufficiently
close to $v_{0}$, and all $g_{n}\left(t\right)$ sufficiently close
to $g_{\ast}$. This corresponds to the case that {[}cf. Eqs.~(\ref{eq:t-sl})
and (\ref{eq:tpar-in-I2}){]} 
\begin{equation}
t_{\parallel,2}<t_{s,\ell}=v_{\ast}/\hat{B}.\label{eq:cond-for-alpha2-in-equilibrium-1}
\end{equation}
We claim that in equilibrium this is equivalent to
\begin{equation}
\left(\frac{1}{\hat{B}}-\frac{1}{B}\right)^{-1}\left(\tau+\theta\right)<v_{\ast}.\label{eq:cond-for-alpha2-in-equilibrium-2}
\end{equation}
To see that, first observe that $g^{\prime}\left(t\right)\equiv0$
in the equilibrium. Therefore, from Eq.~(\ref{eq:alpha2}), we obtain
\begin{equation}
g_{1,\ast}^{\prime}=-\hat{B}\tau.\label{eq:gp1ast-in-equilibrium}
\end{equation}
Substituting this into (\ref{eq:tpar-in-I2}) yields
\begin{equation}
t_{\parallel,2}=\frac{B\left(\tau+\theta\right)}{B-\hat{B}}.\label{eq:tpar2-in-equilibrium}
\end{equation}
Using (\ref{eq:tpar2-in-equilibrium}) and (\ref{eq:cond-for-alpha2-in-equilibrium-1})
we obtain (\ref{eq:cond-for-alpha2-in-equilibrium-2}). 

In the following we follow the notation used by Wilson \cite{wilson_analysis_2001}
and consider a function $F\left(\phi,\psi,\chi\right)$ defined such
that 
\begin{equation}
F\left(g_{n}\left(t\right),v_{n}\left(t\right),v_{n-1}\left(t\right)\right):=v_{n}\left(t+\tau\right),\label{eq:def-F}
\end{equation}
where $v_{n}\left(t+\tau\right)$ is computed by (\ref{eq:vn-t1}).
Note that $F$ still describes a \emph{Gipps-like} model in the sense
of Wilson {[}2001, p.3{]} since we have
\[
\partial_{\phi}F>0,\ \partial_{\psi}F<0,\text{ and }\partial_{\chi}F>0.
\]
This is readily checked by substituting (\ref{eq:alpha2}) into (\ref{eq:vn-t1}).
In contrast to the original Gipps model, (\ref{eq:def-F}) has the
form
\begin{equation}
F\left(\phi,\psi,\chi\right)=\chi+H(\phi,\chi-\psi),\label{eq:F-as-function-of-H}
\end{equation}
with a function 
\begin{equation}
H\left(\phi,\chi-\psi\right):=-\left(g_{1,\ast}^{\prime}\right)_{\mid g_{0}=\phi,\,g_{0}^{\prime}=\chi-\psi}-\tau\hat{B},\label{eq:def-H}
\end{equation}
depending merely on the gap $g_{0}=\phi$ and the speed difference
$g_{0}^{\prime}=\chi-\psi$ of the vehicles {[}cf. Eqn.~(\ref{eq:alpha2})
and (\ref{eq:vn-t1}){]}. As a consequence the stationary gap $g_{\ast}$,
which is approached when following a leader vehicle maintaining a
constant speed $v_{0}$, is \emph{independent of the speed}. Indeed,
$g_{\ast}$ is obtained as a solution of
\begin{equation}
v_{\ast}=F\left(g_{\ast},v_{\ast},v_{\ast}\right)=v_{\ast}+H(g_{\ast},0),\label{eq:equilibrium-condition}
\end{equation}
i.e., $H\left(g_{\ast},0\right)=0$ implicitly defines $g_{\ast}$
independently of the value of the stationary speed $v_{\ast}$ {[}as
long as (\ref{eq:cond-for-alpha2-in-equilibrium-2}) holds{]}. We
obtain 
\begin{equation}
g_{\ast}=\frac{\left(\tau+\theta\right)^{2}}{2}\left(\frac{1}{\hat{B}}-\frac{1}{B}\right)^{-1}.\label{eq:stationary-gap}
\end{equation}
The independence of the stationary speed of the stationary gap has
important consequences. Firstly, an equilibrium solution with $v_{\ast}<v_{{\rm max}}$
only exists in systems with a specific density of vehicles
\begin{equation}
\varrho_{\ast}=\left(g_{\ast}+l+g_{{\rm stop}}\right)^{-1},\label{eq:stationary-density}
\end{equation}
which allows to select uniform gaps of the exact magnitude $g_{\ast}$.
In this case the homogeneous state can at most by neutrally stable
since perturbations, which increase the speed of all vehicles homogeneously
and simultaneously will persist. For lower densities $\varrho<\varrho_{\ast}$
the stationary solution takes the form $v_{\ast}=v_{{\rm max}}$ and
$g_{n}\equiv\varrho^{-1}-l-g_{{\rm stop}}>g_{\ast}$ and is neutrally
stable with respect to small perturbations in the gaps, i.e., also
inhomogeneous gaps $g_{n}$ are stationary with speed $v_{\ast}=v_{{\rm max}}$
as long as $g_{n}>g_{\ast}$. For higher densities $\varrho>\varrho_{\ast}$
and uniform gaps, i.e. $g_{n}\equiv g<g_{\ast}$, there exists no
stationary speed $v_{\ast}$ satisfying (\ref{eq:equilibrium-condition}).
Thus, for any equilibrium solution with gaps $g<g_{\ast}$ the corresponding
equilibrium speed $v_{\ast}$ must violate (\ref{eq:cond-for-alpha2-in-equilibrium-2}),
thus it is simply the equilibrium speed for the original Gipps model,
given as {[}see Wilson, Eqn (3.14){]}
\begin{equation}
v_{\ast}=\left(\frac{1}{\hat{B}}-\frac{1}{B}\right)^{-1}\left(\tau+\theta\right)\left(1-\sqrt{1-\frac{g}{g_{\ast}}}\right).\label{eq:equilibrium-speed-gipps}
\end{equation}
\begin{figure}
\begin{centering}
\includegraphics[width=0.8\textwidth]{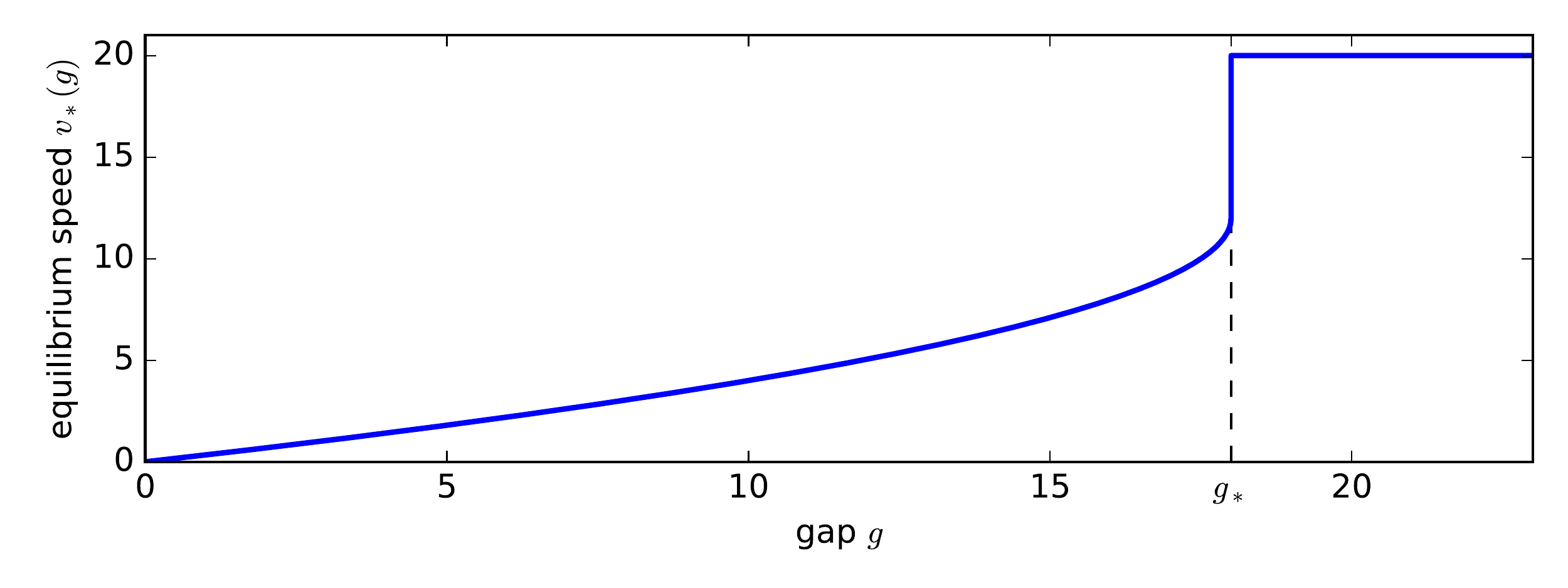}
\par\end{centering}
\caption{\label{fig:equilibrium-speed}Dependence of the equilibrium speed
$v_{\ast}$ on the uniform gap $g$ (speed-headway function). The
dashed line indicates the critical value $g=g_{\ast}$, beyond which
the mapping (\ref{eq:vn-t1}) is active and the vehicles go with maximum
speed in the equilibrium flow. The model parameters for the illustration
are the same as in Fig.~\ref{fig:hypothetical-trajectory-types},
supplemented with $v_{{\rm max}}=20{\rm m/s}$.}

\end{figure}
Figure~\ref{fig:equilibrium-speed}(a) shows the speed-headway function
obtained from this reasoning. Note that it is the natural extension
for the speed-headway function given by Wilson into the region defined
by (\ref{eq:cond-for-alpha2-in-equilibrium-2}). It resolves the ambiguity
of the original Gipps model in this region, which allowed for two
different solutions. Interestingly, the function is discontinuous
in $g_{\ast}$. Whether this is less ``unphysical'' than the multi-valued
function encountered by Wilson is to be discussed. Punzo and Tripodi
proposed speed-headway-density relations based on the original Gipps
model, which resemble the shape but disregard the discontinuity \cite{Punzo2007}.
{[}\textcolor{red}{Optinally, we could elaborate on this a little}{]}

\section{Stability of the Equilibrium Flow\label{sec:Stability-of-the-equilibrium}}

In the following we consider a circular track of length
\begin{equation}
L_{N}^{\ast}=\frac{N}{\varrho_{\ast}},\label{eq:Last}
\end{equation}
with $\varrho_{\ast}$ being the stationary density defined in (\ref{eq:stationary-density}),
i.e., the track allows for equilibrium flows with stationary speed
$v_{\ast}<v_{{\rm max}}$ still fulfilling (\ref{eq:cond-for-alpha2-in-equilibrium-2}).
This should be seen as an auxiliary construction to estimate the wave
propagation within an array of vehicles traveling behind a slower
leading vehicle with $v_{\ell}<v_{{\rm max}}$. In such a situation,
the density $\varrho_{\ast}$ will be approached locally within the
array while traveling at lower speeds. Perturbations induced by irregularities
of the leader's trajectory will be magnified or damped according to
the stability obtained for the equilibrium on the circular track.

To assess the systems stability, we study the system
\[
\left(\begin{array}{c}
v_{n}(t)\\
g_{n}(t)
\end{array}\right)\mapsto\left(\begin{array}{c}
v_{n}(t+\tau)\\
g_{n}(t+\tau)
\end{array}\right)=\left(\begin{array}{c}
F\left(g_{n}\left(t\right),v_{n}\left(t\right),v_{n-1}\left(t\right)\right)\\
G\left(g_{n}(t),v_{n}(t),g_{n-1}(t),v_{n-1}(t),v_{n-2}(t)\right)
\end{array}\right),
\]
where all indices should be considered modulo $N$ and
\begin{align*}
G\left(g_{n},v_{n},g_{n-1},v_{n-1},v_{n-2}\right) & =g_{n}+\frac{\tau}{2}\left(v_{n-1}-v_{n}+F\left(g_{n-1},v_{n-1},v_{n-2}\right)-F\left(g_{n},v_{n},v_{n-1}\right)\right)
\end{align*}
describes the new front gap $g_{n}\left(t+\tau\right)$ after the
$n$-th vehicle has traveled for a time $\tau$ with constant acceleration
$a_{n}=\left(v_{n}\left(t+\tau\right)-v_{n}\left(t\right)\right)/\tau$.
The linearization of this map in a neighborhood of the equilibrium
flow with $v_{n}=v_{\ast}+\delta_{n}$, $g_{n}=g_{\ast}+\eta_{n}$,
yields
\begin{align}
F\left(g_{n},v_{n},v_{n-1}\right) & \approx v_{\ast}+DF_{\ast}\left(\eta_{n},\delta_{n},\delta_{n-1}\right)\nonumber \\
 & =v_{\ast}+\partial_{\phi}F_{\ast}\eta_{n}+\partial_{\psi}F_{\ast}\delta_{n}+\partial_{\chi}F_{\ast}\delta_{n-1},\label{eq:linearization-F}\\
G\left(g_{n},v_{n},g_{n-1},v_{n-1},v_{n-2}\right) & \approx g_{\ast}+DG_{\ast}\left(\eta_{n},\delta_{n},\eta_{n-1},\delta_{n-1},\delta_{n-2}\right)\nonumber \\
 & =g_{\ast}+\left(1-\frac{\tau}{2}\partial_{\phi}F_{\ast}\right)\eta_{n}-\frac{\tau}{2}\left(1+\partial_{\psi}F_{\ast}\right)\delta_{n}+\frac{\tau}{2}\partial_{\phi}F_{\ast}\eta_{n-1}\nonumber \\
 & +\frac{\tau}{2}\left(1+\partial_{\psi}F_{\ast}-\partial_{\chi}F_{\ast}\right)\delta_{n-1}+\frac{\tau}{2}\partial_{\chi}F_{\ast}\delta_{n-2},\label{eq:linearization-G}
\end{align}
with $\partial_{j}F_{\ast}=\partial_{j}F\left(g_{\ast},v_{\ast},v_{\ast}\right),$
$j=\phi,\psi,\chi$. Next, we calculate the derivatives of $F\left(\phi,\psi,\chi\right)$.
For notational convenience, we introduce a function
\[
D\left(\phi,\chi-\psi\right):=\left(B-\hat{B}\right)^{2}\tau^{2}+4\left(B-\hat{B}\right)\left(\left(\tau\theta+\theta^{2}\right)B+\left(\chi-\psi\right)\tau+2\phi\right).
\]
This gives {[}cf. Eqn.~(\ref{eq:slns-gp1}){]}
\begin{equation}
g_{1,\ast}^{\prime}=\frac{1}{2}\left(B-\hat{B}\right)\tau+B\theta-\frac{1}{2}\sqrt{D\left(\phi,\chi-\psi\right)}.\label{eq:gp1ast-as-function-of-D}
\end{equation}
Further note that from (\ref{eq:gp1ast-in-equilibrium}) it follows
that in equilibrium we have
\begin{equation}
\sqrt{D\left(g_{\ast},0\right)}=B\left(\tau+2\theta\right)+\hat{B}\tau.\label{eq:D-in-equilibrium}
\end{equation}
Refering to Eqns.~(\ref{eq:F-as-function-of-H}),(\ref{eq:def-H}),
and (\ref{eq:gp1ast-as-function-of-D}), we calculate
\[
\partial_{j}F\left(\phi,\psi,\chi\right)=\partial_{j}\chi+\frac{1}{4}\sqrt{D\left(\phi,\chi-\psi\right)}^{-1}\partial_{j}D\left(\phi,\chi-\psi\right),\ j=\phi,\psi,\chi.
\]
Thus, in equilibrium Eqn.~(\ref{eq:D-in-equilibrium}) gives
\[
\partial_{j}F_{\ast}=\partial_{j}\chi+\frac{\partial_{j}D}{4\left(B\left(\tau+2\theta\right)+\hat{B}\tau\right)},
\]
where $\partial_{j}D=4\left(B-\hat{B}\right)\partial_{j}\left(\left(\chi-\psi\right)\tau+2\phi\right)$.
Defining
\begin{equation}
\xi:=\frac{B-\hat{B}}{B\left(\tau+2\theta\right)+\hat{B}\tau}>0,\label{def-xi}
\end{equation}
we find that
\begin{equation}
\partial_{\phi}F_{\ast}=2\xi,\ \partial_{\psi}F_{\ast}=-\tau\xi,\text{ and }\partial_{\chi}F_{\ast}=1+\tau\xi.\label{eq:DFphi-psi-chi-as-function-of-xi}
\end{equation}
Thus, (\ref{eq:linearization-F}) and (\ref{eq:linearization-G})
may be written as
\begin{align}
DF_{\ast}\left(\eta_{n},\delta_{n},\delta_{n-1}\right) & =2\xi\eta_{n}-\tau\xi\delta_{n}+\left(1+\tau\xi\right)\delta_{n-1}\label{eq:DF-as-a-function-of-xi}\\
DG_{\ast}\left(\eta_{n},\delta_{n},\eta_{n-1},\delta_{n-1},\delta_{n-2}\right) & =\left(1-\tau\xi\right)\eta_{n}-\frac{\tau}{2}\left(1-\tau\xi\right)\delta_{n}+\tau\xi\eta_{n-1}-\tau^{2}\xi\delta_{n-1}+\frac{\tau}{2}\left(1+\tau\xi\right)\delta_{n-2}\label{eq:DG-as-a-function-of-xi}
\end{align}
To determine the stability we determine characteristic multipliers
$\mu$ for the linearized system. That is solutions to
\begin{align}
DF_{\ast}\left(\eta_{n},\delta_{n},\delta_{n-1}\right) & =\mu\delta_{n}\label{eq:eigenvec-DF-comp}\\
DG_{\ast}\left(\eta_{n},\delta_{n},\eta_{n-1},\delta_{n-1},\delta_{n-2}\right) & =\mu\eta_{n}.\label{eq:eigenvec-DG-comp}
\end{align}
The rotational symmetry of the systems suggests a rotating wave ansatz
for the eigenmodes, i.e., 
\begin{align}
\left(\begin{array}{c}
\delta_{n}\\
\eta_{n}
\end{array}\right) & =\omega_{k}^{n}\left(\begin{array}{c}
\delta\\
\eta
\end{array}\right),\label{eq:rotating-wave}
\end{align}
with an $N$-th root of unity 
\[
\omega_{k}=\exp\left(\frac{2\pi k}{N}\cdot i\right),
\]
where $i=\sqrt{-1}$ and $\eta,\delta\in\mathbb{C}$ are arbitrary
complex numbers with $\eta\cdot\delta\ne0$. Let us first consider
the case that $\eta=0$, which corresponds to unperturbed gaps. From
(\ref{eq:DG-as-a-function-of-xi}) and (\ref{eq:eigenvec-DG-comp})
we obtain
\begin{equation}
-\frac{\tau}{2}\left(1-\tau\xi\right)-\tau^{2}\xi\omega_{k}^{-1}+\frac{\tau}{2}\left(1+\tau\xi\right)\omega_{k}^{-2}=0.\label{eq:DGeq0}
\end{equation}
Solving (\ref{eq:DGeq0}) for $\omega_{k}^{-1}$ yields
\[
\omega_{k}=\frac{\tau\xi+1}{\tau\xi\pm1},
\]
which implies $k=0$. Equations (\ref{eq:DF-as-a-function-of-xi})
and (\ref{eq:eigenvec-DF-comp}) give the corresponding multiplier
\[
\mu_{0,1}=\left(1+\tau\xi\right)\omega_{k}^{-1}-\tau\xi=1.
\]
As argued in Sec.~\ref{sec:The-equilibrium-flow} we obtain a neutrally
stable mode corresponding a uniform variation of all vehicles' speeds.

\begin{figure}
\centering{}\hspace{-1cm}(a)%
\begin{minipage}[t][1\totalheight][b]{0.55\columnwidth}%
\begin{center}
\includegraphics[width=1\textwidth]{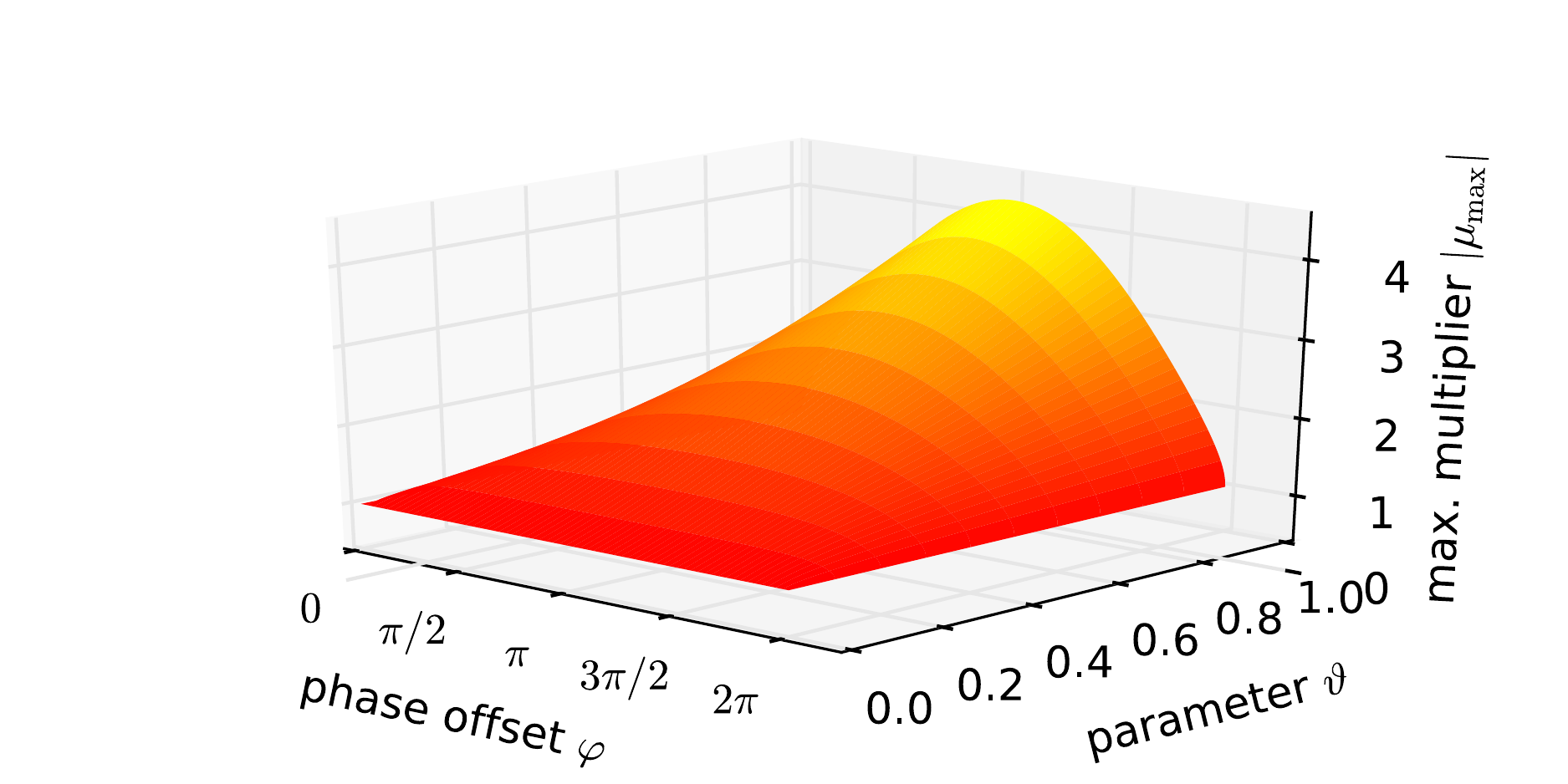}
\par\end{center}%
\end{minipage}~~~~(b)%
\begin{minipage}[t][1\totalheight][b]{0.44\columnwidth}%
\begin{center}
\includegraphics[width=1\textwidth]{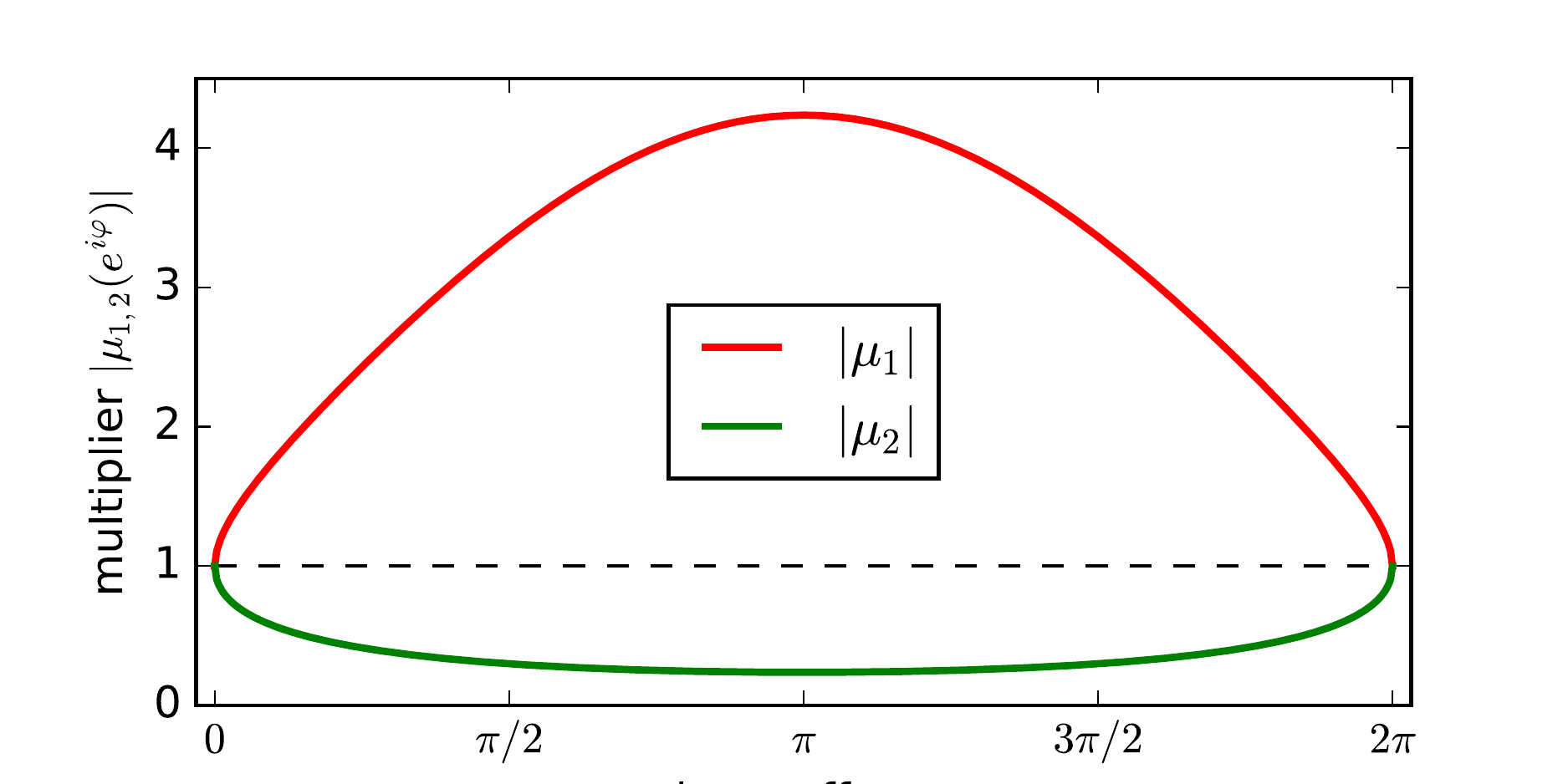}
\par\end{center}%
\end{minipage}\caption{\label{fig:multiplier_equilibrium}Plot~(a): Maximal modulus $\left|\mu_{1}\left(\omega\right)\right|>\left|\mu_{2}\left(\omega\right)\right|$
of the solutions of (\ref{eq:char-eq-1})\textendash (\ref{eq:char-eq-2})
in dependence of the parameter $\vartheta=\tau\xi$ and the angle
$\varphi$ of $\omega=e^{i\varphi}$. The range $\vartheta\in\left[0,1\right]$
contains all reasonable parametrizations of the model, cf.~(\ref{eq:vartheta}).
Plot~(b) shows both multipliers $\mu_{1,2}\left(e^{i\varphi}\right)$
for $\varphi\in\left[0,2\pi\right]$.}
\end{figure}

\begin{figure}
\centering{}(a)%
\begin{minipage}[t][1\totalheight][b]{0.49\columnwidth}%
\begin{center}
\includegraphics[width=1\textwidth]{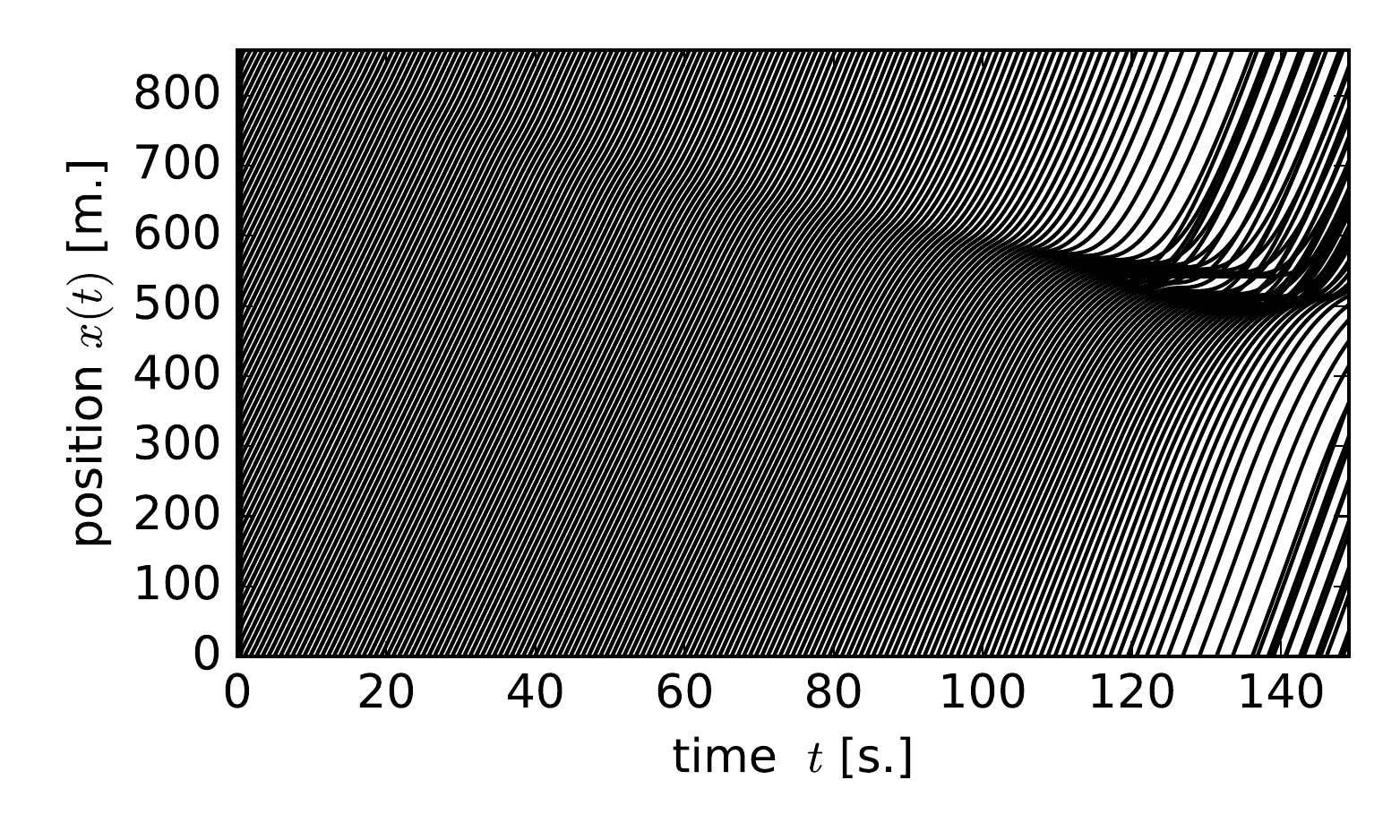}
\par\end{center}%
\end{minipage}(b)%
\begin{minipage}[t][1\totalheight][b]{0.49\columnwidth}%
\begin{center}
\includegraphics[width=1\textwidth]{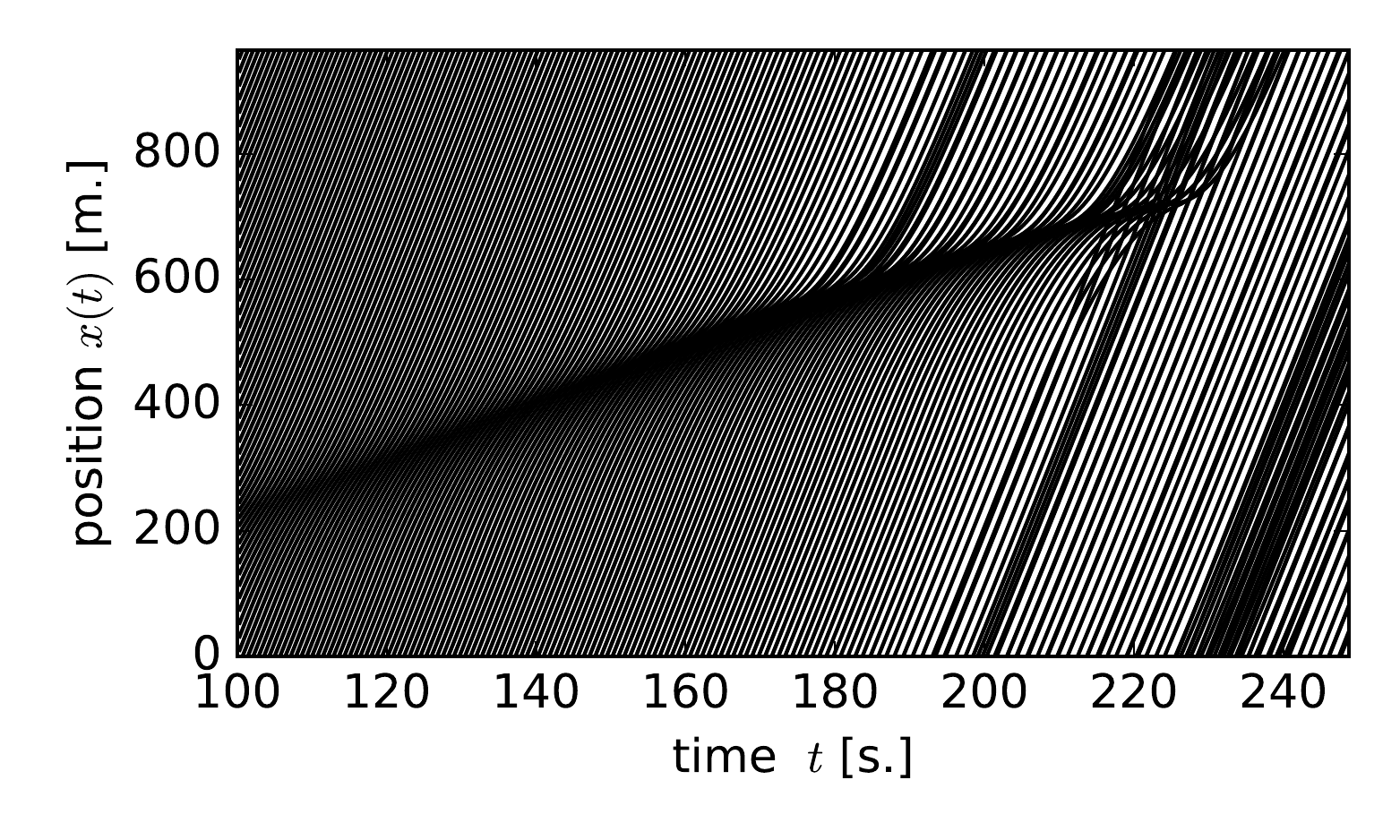}
\par\end{center}%
\end{minipage}\caption{\label{fig:collsions-circular-track}Collisions on a circular track
with $N=50$ vehicles with initial speed $v_{0}=v_{\ast}+1\approx21.8{\rm m/s}$
for $v_{\ast}=\left(\tau+\theta\right)\left(1/\hat{B}-1/B\right)$,
where $\tau=0.66$, $\theta=0.33$, $B=1.5$, and $\hat{B}=1.4$.
Further, the maximal speed was set to $v_{{\rm max}}=30{\rm m/s}.$
At $t=0.$ the vehicles are distributed uniformly on the track and
one vehicle's initial speed is perturbed by setting it to $\tilde{v}_{0}=0.9\,v_{0}$.
For Plot~(a) the track length is $L=L_{50}^{\ast}\approx865{\rm m}$
{[}see (\ref{eq:Last}){]}. For Plot~(b): $L=L_{50}^{\ast}+100{\rm m}$.}
\end{figure}

For the calculation of the remaining multipliers we may assume that
$\eta\ne0$. Without loss of generality we set $\eta=1$. Then, Eqs.~(\ref{eq:eigenvec-DF-comp})\textendash (\ref{eq:eigenvec-DG-comp})
take the form
\begin{align}
\mu\delta & =2\xi-\tau\xi\delta+\left(1+\tau\xi\right)\delta\omega,\label{eq:char-eq-1}\\
\mu & =\left(1-\tau\xi\right)-\frac{\tau}{2}\left(1-\tau\xi\right)\delta+\tau\xi\omega-\tau^{2}\xi\delta\omega+\frac{\tau}{2}\left(1+\tau\xi\right)\delta\omega^{2},\label{eq:char-eq-2}
\end{align}
where we let $\omega=\omega_{k}^{-1}$. Eqn.~(\ref{eq:char-eq-1})
yields
\[
\delta=-\frac{2\,\xi}{\left(\omega-1\right)\tau\xi-\mu+\omega}.
\]
Using this in (\ref{eq:char-eq-2}) gives:
\begin{equation}
\mu_{1,2}=h\left(\omega,\tau\xi\right)\pm\sqrt{h\left(\omega,\tau\xi\right)^{2}-\omega},\label{eq:mu12}
\end{equation}
with $h\left(\vartheta,\omega\right)=\left(\omega-1\right)\vartheta+\frac{\omega+1}{2}$.
Note that $\mu$ and thus the stability only depends on the product
\begin{equation}
\vartheta=\tau\xi=\frac{B-\hat{B}}{B\left(1+2\theta/\tau\right)+\hat{B}}.\label{eq:vartheta}
\end{equation}
Figure~\ref{fig:multiplier_equilibrium}(a) shows the larger of the
two moduli $\left|\mu_{1}\right|$ and $\left|\mu_{2}\right|$ given
in dependence of $\vartheta$ and the argument $\varphi$ of $\omega=e^{i\varphi}$.
The numerical results support the conjecture that the maximal multiplier
is obtained for $\varphi=\pi$ independently of the value of $\vartheta$.
An analogous observation was presented by Wilson for the case $B\le\hat{B}$.
We remind that the corresponding instability is a short wavelength
instability, see \cite{wilson_analysis_2001,bando_dynamical_1995}.
Figure~\ref{fig:multiplier_equilibrium}(b) shows both curves $\varphi\mapsto\left|\mu_{j}\left(\exp\left(i\varphi\right)\right)\right|$,
$j=1,2$, for $\vartheta=1.0$. Similarly, for all values $\vartheta>0$
we obtain a branch $\left|\mu_{1}\right|>1$. Hence the equilibrium
flow for $B>\hat{B}$ is unstable for all choices of parameters as
long as Eqn.~(\ref{eq:cond-for-alpha2-in-equilibrium-2}) is fulfilled.
Indeed, numerical simulations suggest that collisions are unavoidable,
which is not too surprising since all drivers systematically underestimate
their leader's disposition to brake hard. Figure~\ref{fig:collsions-circular-track}
shows two scenarios. In Panel~(a) the track length is set exactly
to $L=L_{N}^{\ast}$ {[}see (\ref{eq:Last}){]} leading to an earlier
collision. In Panel~(b) the track length was increased such that
each inter-vehicle gap increased by $2{\rm m}$. Interestingly, the
direction of the wave emerging from the perturbation and ultimately
leading to a collision propagates downstream, here. This stands in
contrast to the wave in (a), which travels downstream. 

\section{Simulation Experiments\label{sec:Simulation-Experiments}}

\begin{figure}
\begin{centering}
(a)%
\begin{minipage}[t][1\totalheight][b]{0.3\columnwidth}%
\begin{center}
\includegraphics[width=1\textwidth]{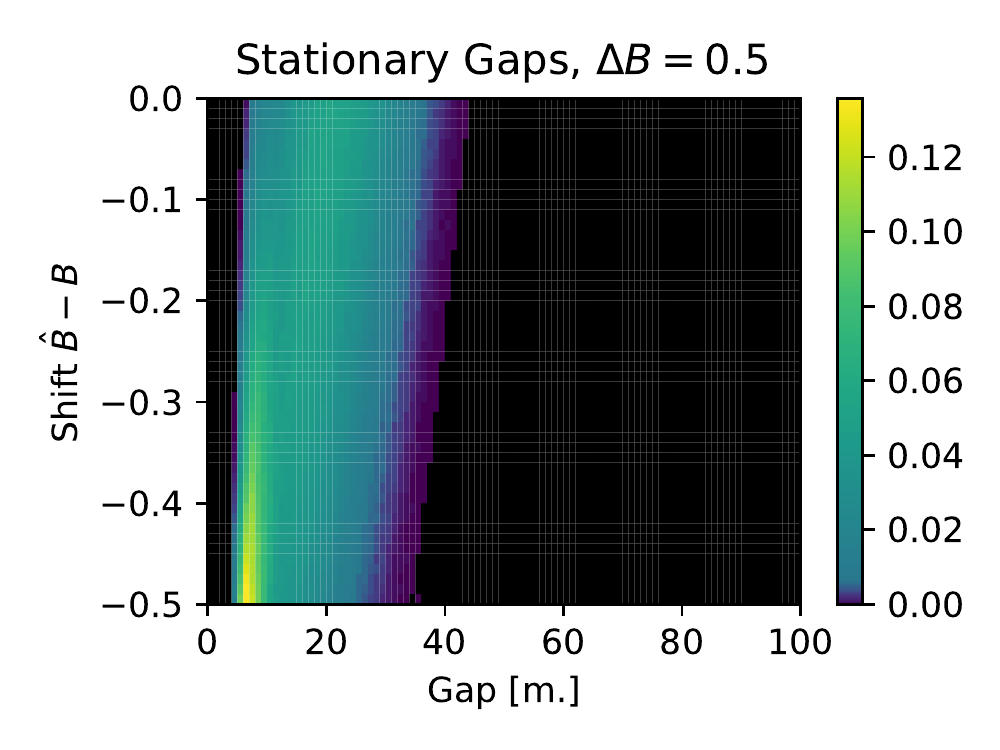}
\par\end{center}%
\end{minipage}(b)%
\begin{minipage}[t][1\totalheight][b]{0.3\columnwidth}%
\begin{center}
\includegraphics[width=1\textwidth]{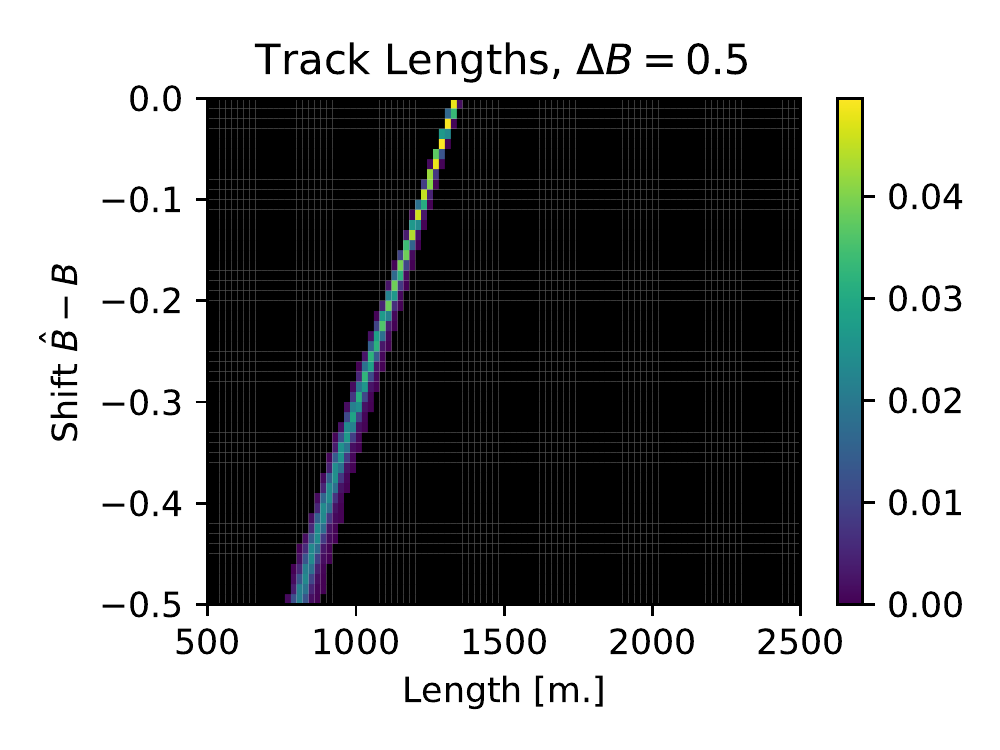}
\par\end{center}%
\end{minipage}(c)%
\begin{minipage}[t][1\totalheight][b]{0.3\columnwidth}%
\begin{center}
\includegraphics[width=1\textwidth]{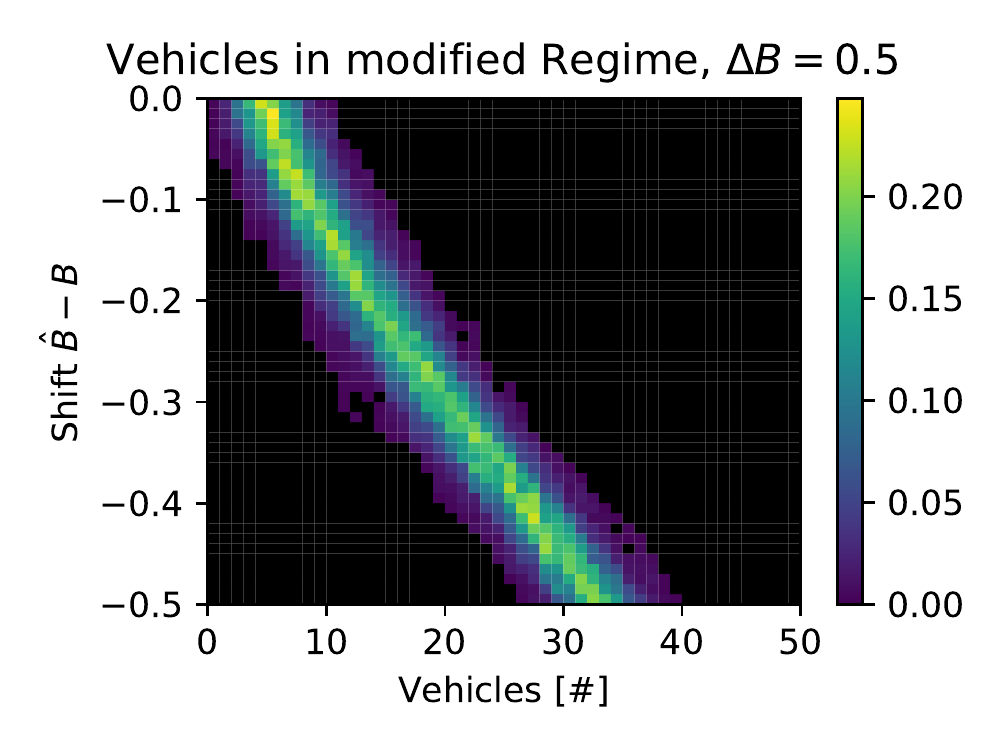}
\par\end{center}%
\end{minipage}
\par\end{centering}
\begin{centering}
(d)%
\begin{minipage}[t][1\totalheight][b]{0.3\columnwidth}%
\begin{center}
\includegraphics[width=1\textwidth]{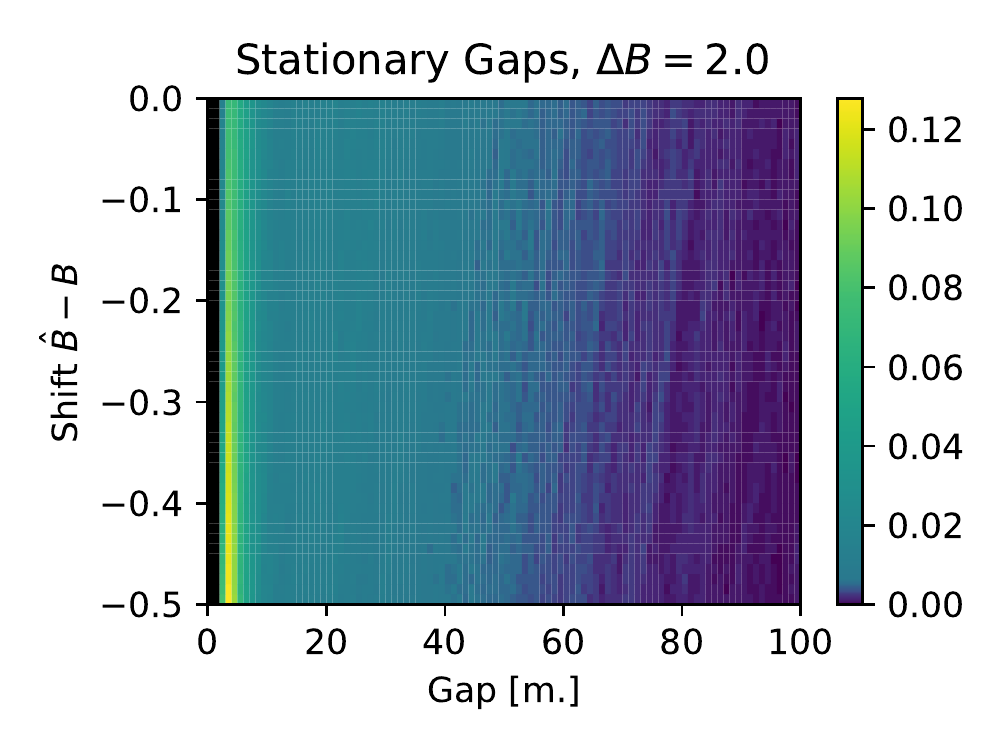}
\par\end{center}%
\end{minipage}(e)%
\begin{minipage}[t][1\totalheight][b]{0.3\columnwidth}%
\begin{center}
\includegraphics[width=1\textwidth]{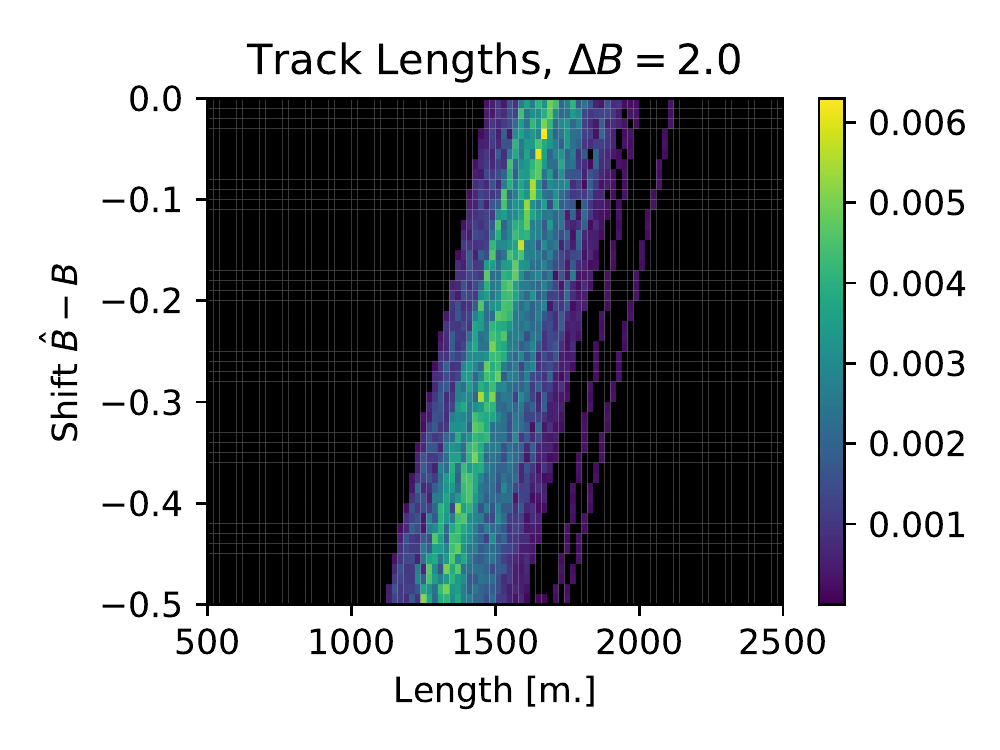}
\par\end{center}%
\end{minipage}(f)%
\begin{minipage}[t][1\totalheight][b]{0.3\columnwidth}%
\begin{center}
\includegraphics[width=1\textwidth]{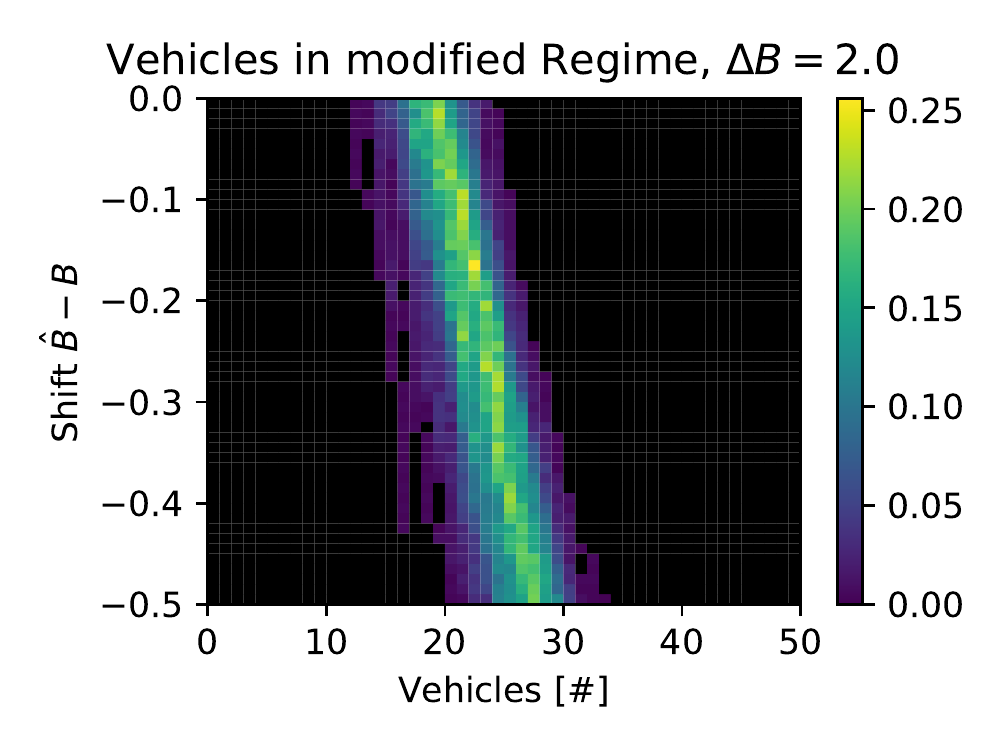}
\par\end{center}%
\end{minipage}
\par\end{centering}
\caption{\label{fig:initial-config-distributions-GippsX}Distributions of different
quantities related to the equilibrium flow in a heterogeneous ensemble
of vehicles with stationary speed $v_{\ast}=20m/s$ for the model
extension described in Section~\ref{sec:vsafe-for-b-ge-bhat}. The
distributions obtained for different values of the bias $\Delta\hat{B}=\hat{B}_{i}-B_{i}$
are shown as pixel rows and for two values of $\Delta B$ ((a)\textendash (c):
$\Delta B=0.5m/s^{2}$, (d)\textendash (f): $\Delta B=2.0m/s^{2}$,
for both cases $\bar{B}=3.0m/s^{2}$). (a) and (d): Stationary gaps
(including $g_{{\rm stop}}=2.0m.$); (b) and (e): Track lengths, cf.
(\ref{eq:B-dependent-track-length}); (c) and (f): Number of vehicles
for which the formula (\ref{eq:stationary-gap}) is employed to determine
the stationary gap.}
\end{figure}

In this section we present some simulation results on a circular track
for a set of $50$ non-identical vehicles. In particular, we are interested
in the situation where $B>\hat{B}$ may arise without an assumption
of underestimating $\hat{B}$. To this end we assumed that the values
$B_{i}$, $i=1,...,50$, for the different vehicles are uniformly
distributed on an interval $[\bar{B}-\Delta B,\bar{B}+\Delta B]$.
Further we varied the expected leader deceleration rates $\hat{B}_{i}=B_{i}+\Delta\hat{B}$
by means of a constant, vehicle-independent bias $\Delta\hat{B}$.
For a simulation with a particular realization of $\boldsymbol{B}=(B_{1},...,B_{50})$
we assumed the length of the circular track to be equal to
\begin{equation}
L(\boldsymbol{B})=50\cdot(l+g_{{\rm stop}})+\sum_{i=1}^{50}g_{i,i+1}^{\ast}(v_{\ast},B_{i},\hat{B}_{i+1}),\label{eq:B-dependent-track-length}
\end{equation}
where the stationary gap $g_{i,i+1}^{\ast}$between vehicle $i$ and
$i+1$ is the smallest possible stationary spatial following headway,
see Section~\ref{sec:The-equilibrium-flow}. Figure~\ref{fig:initial-config-distributions-GippsX}
shows the distributions of $L(\boldsymbol{B})$, $g_{i,i+1}^{\ast}$,
and the number of vehicles where $g_{i,i+1}^{\ast}$ is calculated
according to the proposed extension, i.e., from (\ref{eq:stationary-gap}),
and differs from the stationary gap obtained for a conventional Gipps
model. The distributions are normalized histograms generated from
$500$ realizations per value of $\Delta\hat{B}$. 

For the stationary gaps {[}Fig.~\ref{fig:initial-config-distributions-GippsX}(a),(d){]}
the most pronounced feature is the accumulation of gaps at the lower
bound of the distribution's support owing to the relatively small
headways obtained for the case (\ref{eq:stationary-gap}). By definition,
this accumulation is correlated to the number of vehicles, for which
(\ref{eq:stationary-gap}) was employed, see (c) and (f). This is
most clearly seen for the upper array of plots, where $\Delta B=0.5m/s^{2}$
is smaller as this number varies stronger with $\Delta\hat{B}$ than
for the case $\Delta B=2.0m/s^{2}$ shown in the lower array of plots.
The small number of vehicles, for which the stationary gap is computed
according to (\ref{eq:stationary-gap}) for small $\Delta\hat{B}$
{[}cf. (c){]} follows from the fact that for small differences in
$B$ and $\hat{B}$, (\ref{eq:stationary-gap}) is not employed, since
for the hypothesized evolution of the headway $g(t)$, a tangency
to $g=0$ is not possible until the leader has come to a full stop
and the original formula used by Gipps still yields valid solutions.
Although a stronger concentration of smaller headways can be observed
for the case $\Delta B=2.0m/s^{2}$ {[}cf. (a) and (d){]}, the corresponding
distributions of headways have longer tails leading to larger total
sums, which determine the track lengths $L(\boldsymbol{B})$, cf.
(b) and (e), as well as Eqn.~(\ref{eq:B-dependent-track-length}).

\begin{figure}
\begin{centering}
(a)%
\begin{minipage}[t][1\totalheight][b]{0.3\columnwidth}%
\begin{center}
\includegraphics[width=1\textwidth]{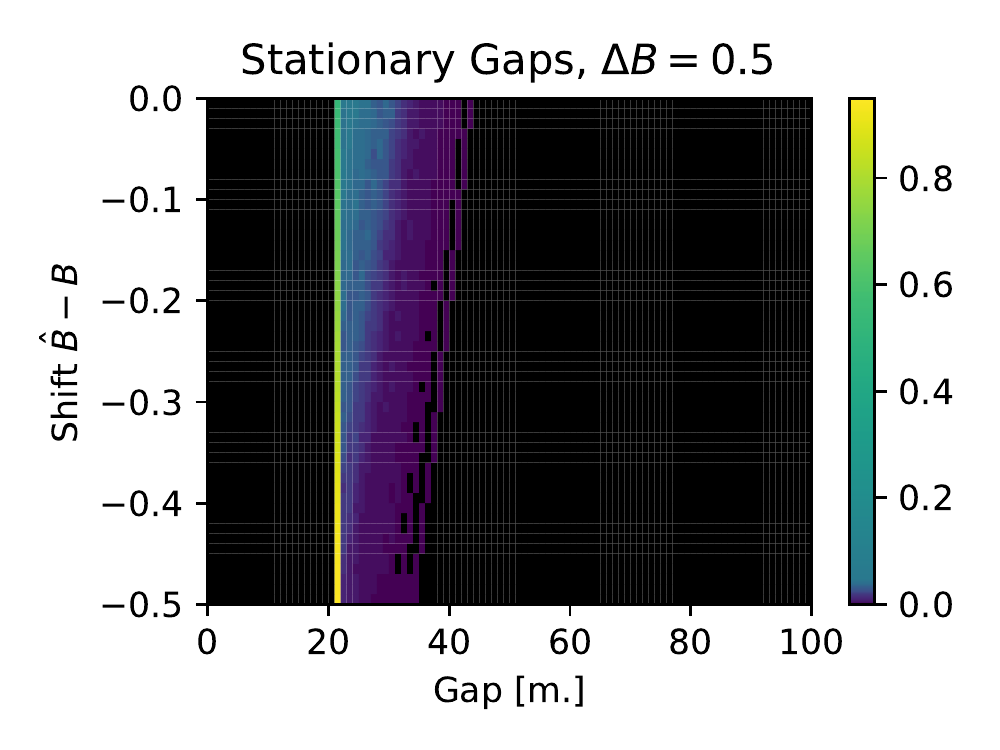}
\par\end{center}%
\end{minipage}(b)%
\begin{minipage}[t][1\totalheight][b]{0.3\columnwidth}%
\begin{center}
\includegraphics[width=1\textwidth]{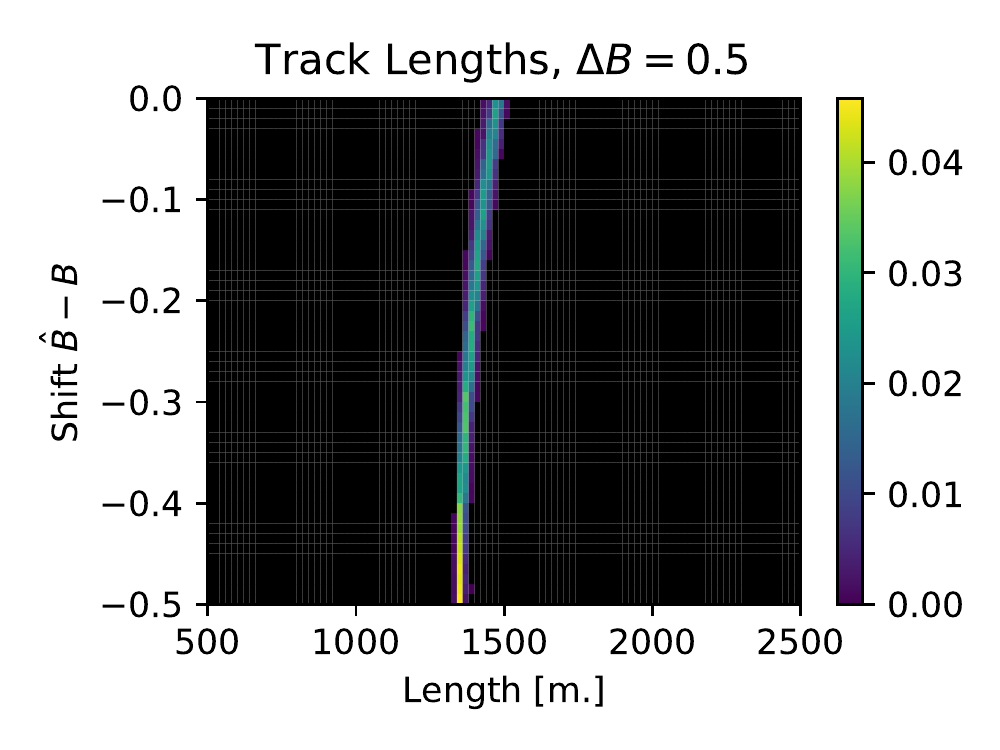}
\par\end{center}%
\end{minipage}(c)%
\begin{minipage}[t][1\totalheight][b]{0.3\columnwidth}%
\begin{center}
\includegraphics[width=1\textwidth]{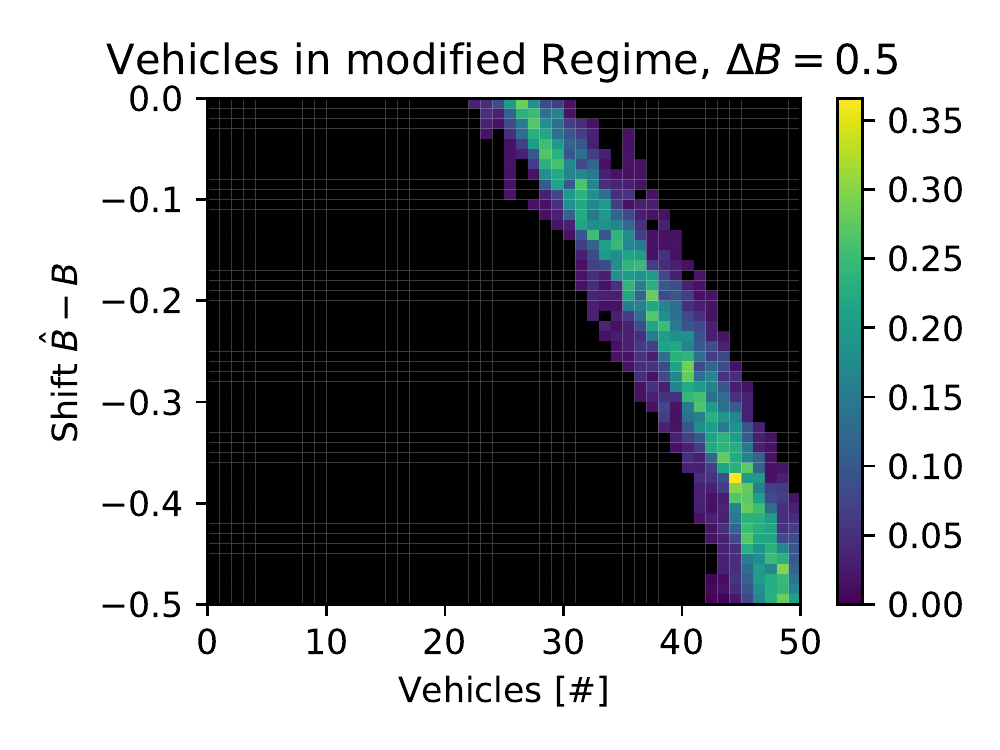}
\par\end{center}%
\end{minipage}
\par\end{centering}
\begin{centering}
(d)%
\begin{minipage}[t][1\totalheight][b]{0.3\columnwidth}%
\begin{center}
\includegraphics[width=1\textwidth]{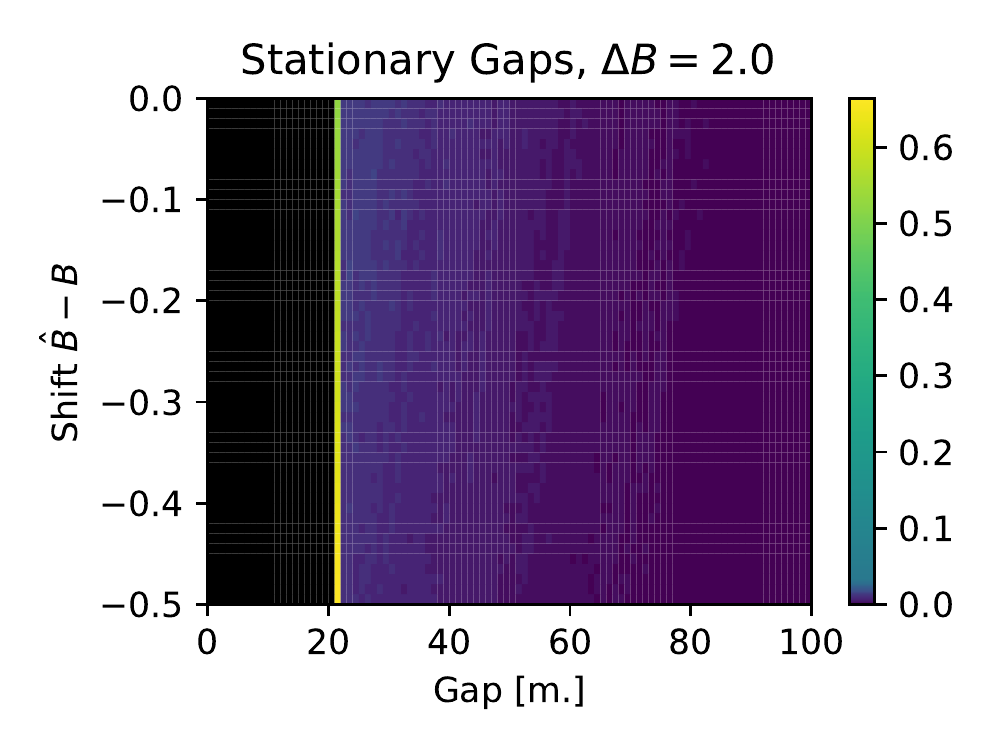}
\par\end{center}%
\end{minipage}(e)%
\begin{minipage}[t][1\totalheight][b]{0.3\columnwidth}%
\begin{center}
\includegraphics[width=1\textwidth]{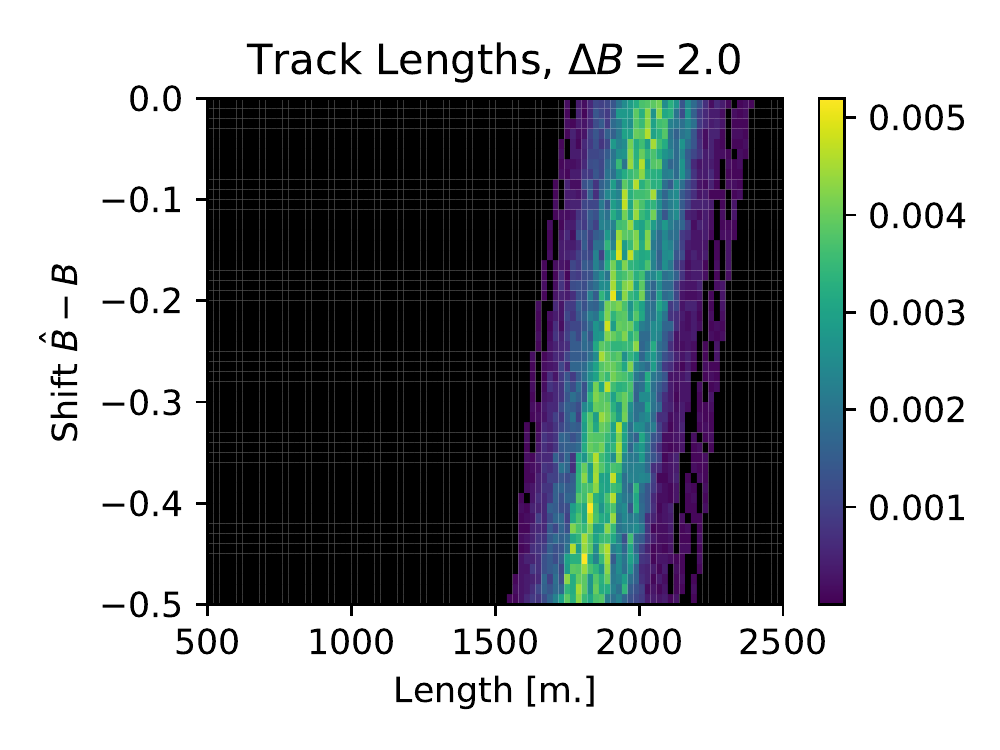}
\par\end{center}%
\end{minipage}(f)%
\begin{minipage}[t][1\totalheight][b]{0.3\columnwidth}%
\begin{center}
\includegraphics[width=1\textwidth]{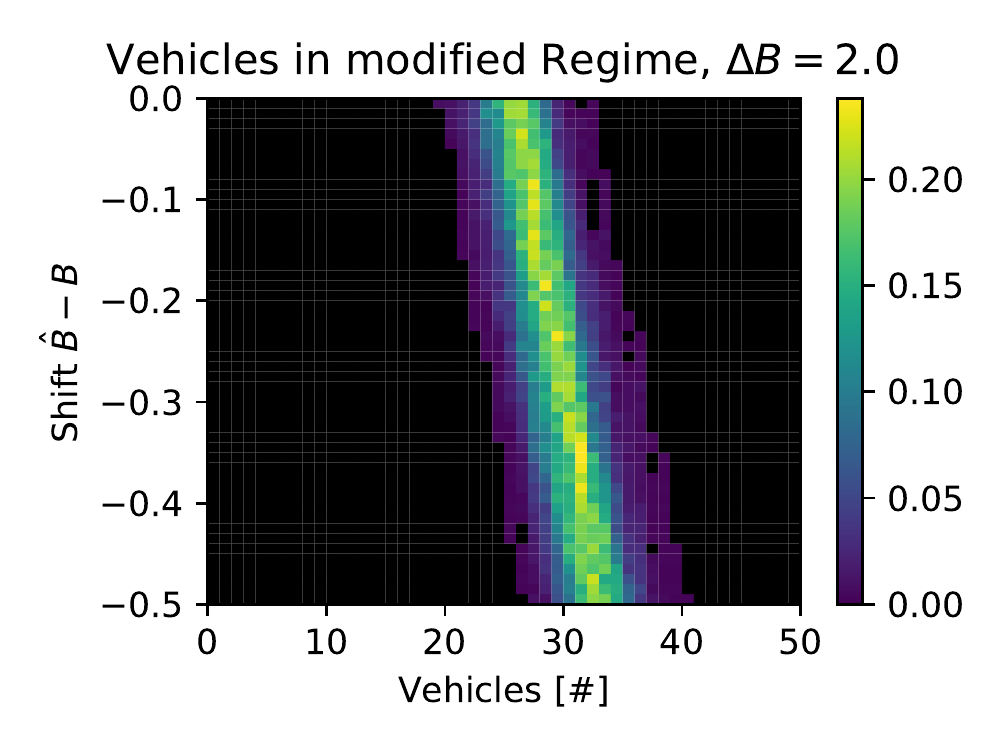}
\par\end{center}%
\end{minipage}
\par\end{centering}
\caption{\label{fig:initial-config-distributions-Krauss}Distributions of different
quantities related to the equilibrium flow in a heterogeneous ensemble
of vehicles with stationary speed $v_{\ast}=20m/s$ obtained when
resolving $B>\hat{B}$ as in (\ref{eq:resolution-SUMO}). Parameters:
$\bar{B}=3.0m/s^{2}$, (a)\textendash (c): $\Delta B=0.5m/s^{2}$,
(d)\textendash (f): $\Delta B=2.0m/s^{2}$, cf. Fig.~\ref{fig:initial-config-distributions-GippsX}.}
\end{figure}

Figure~\ref{fig:initial-config-distributions-Krauss} shows the analogous
results to those depicted in Fig.~\ref{fig:initial-config-distributions-GippsX}
for the case where the situation $B>\hat{B}$ is resolved by assuming
the leader's estimated braking rate to be not less than the followers,
cf. (\ref{eq:resolution-SUMO}). Note that this implies that for each
pair of subsequent vehicles with $B_{{\rm follower}}>B_{{\rm leader}}+\Delta\hat{B}$
this leads to an identical stationary headway of $g_{\ast}=(\tau+\theta)\cdot v_{\ast}$.
This is the the reason for the vertical array of peaks in (a) and
(d). It is clear that the resolution (\ref{eq:resolution-SUMO}) will
always lead to a larger stationary headway than the minimal safe headway
as derived in Section~\ref{sec:vsafe-for-b-ge-bhat}, cf. Fig.~\ref{fig:initial-config-distributions-GippsX}(a),(d).
This is also the reason for the larger total sum of headways observable
in Fig.~\ref{fig:initial-config-distributions-Krauss}(b) and (e)
when compared to Fig.~\ref{fig:initial-config-distributions-GippsX}(b)
and (e).

\begin{figure}
\begin{centering}
(a)%
\begin{minipage}[t][1\totalheight][b]{0.45\columnwidth}%
\begin{center}
\includegraphics[width=1\textwidth]{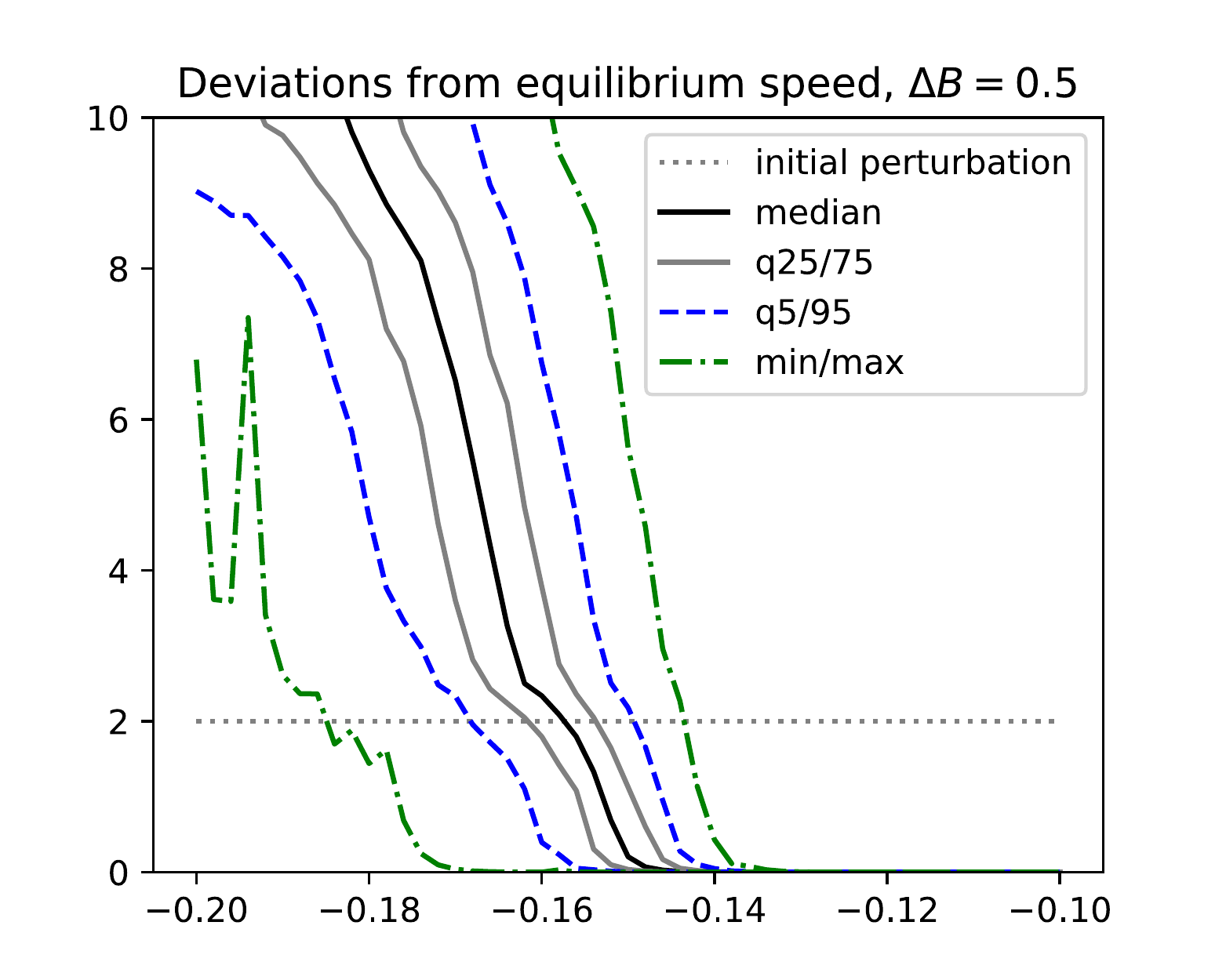}
\par\end{center}%
\end{minipage}(b)%
\begin{minipage}[t][1\totalheight][b]{0.45\columnwidth}%
\begin{center}
\includegraphics[width=1\textwidth]{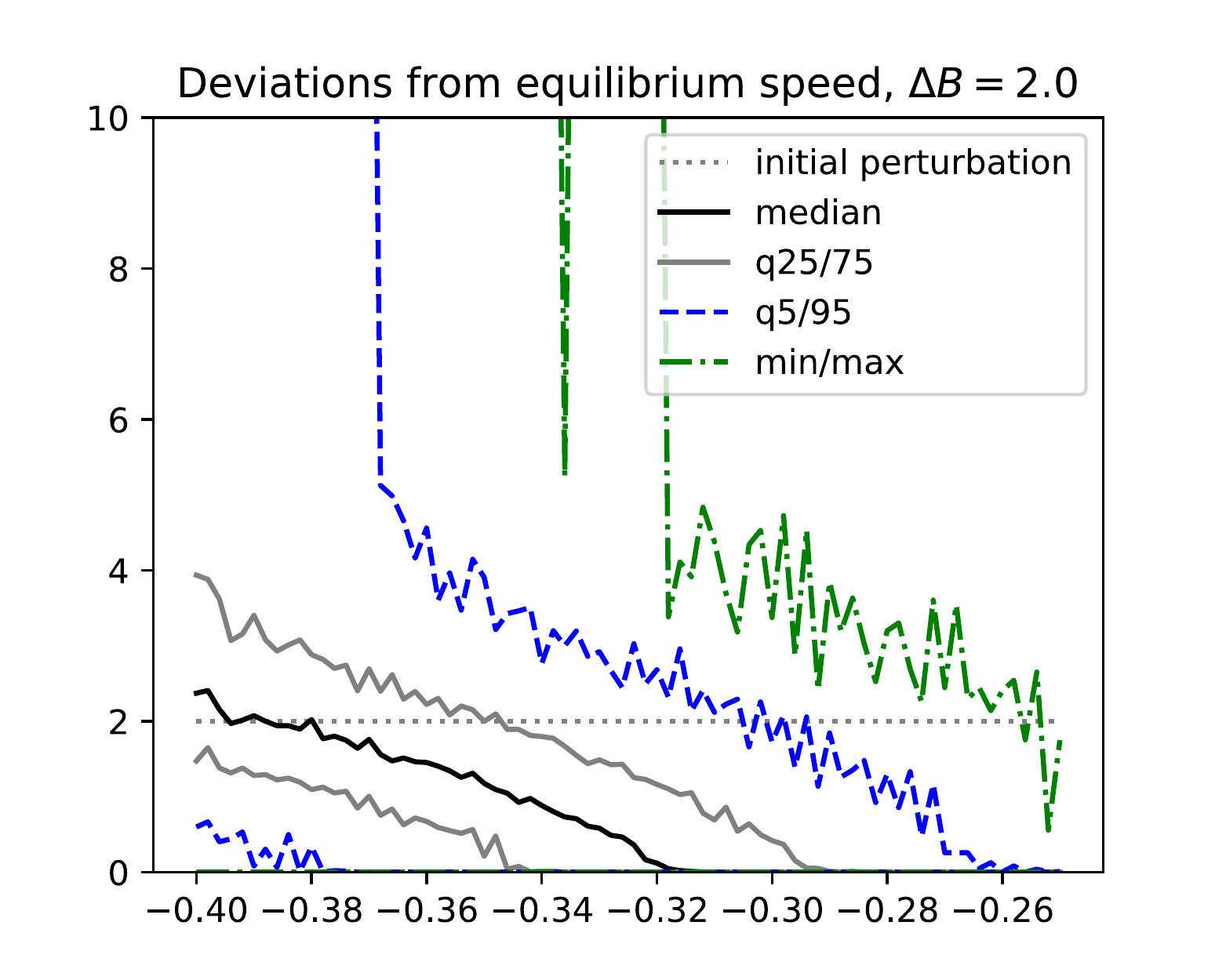}
\par\end{center}%
\end{minipage}
\par\end{centering}
\centering{}(c)%
\begin{minipage}[t][1\totalheight][b]{0.95\columnwidth}%
\begin{center}
\includegraphics[width=1\textwidth]{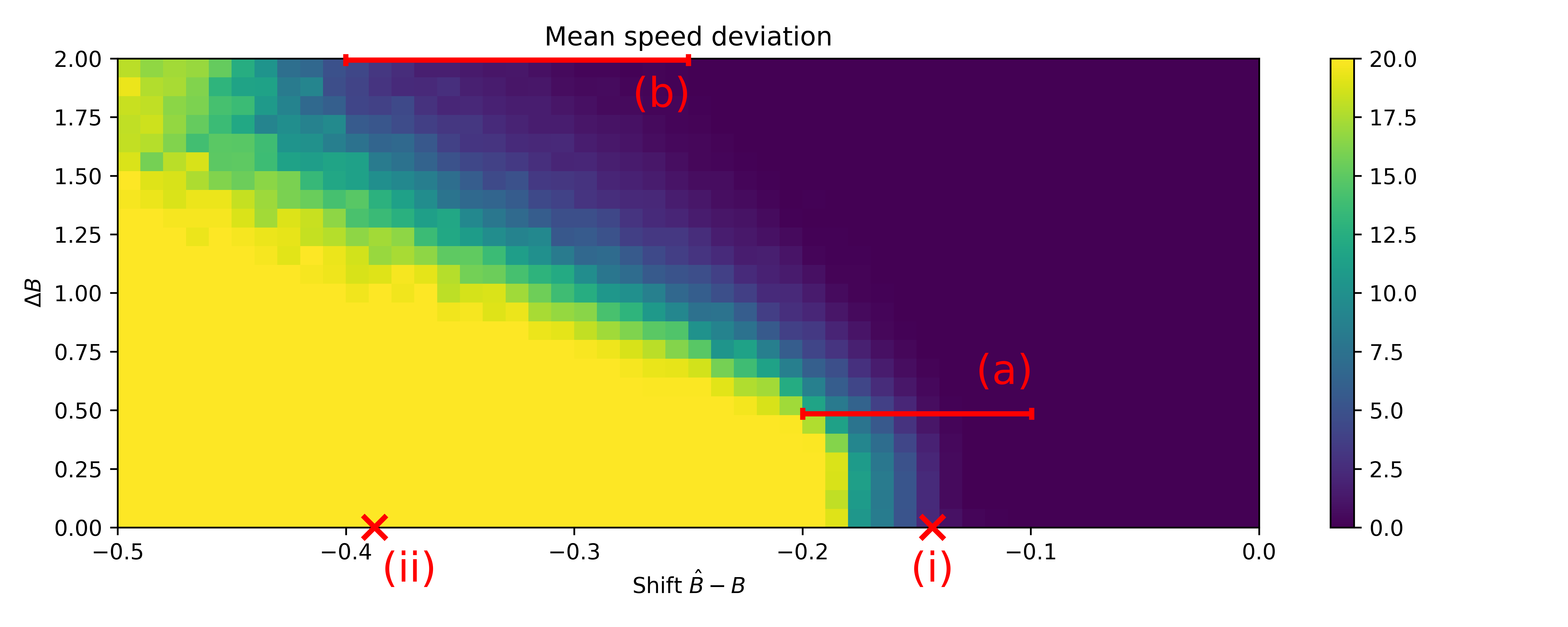}
\par\end{center}%
\end{minipage}\caption{\label{fig:stability-scan}Observed stability of equilibrium flow
for heterogeneous ensembles. (a) and (b): For two fixed values of
the variability $\Delta B$ several percentiles of the observed distribution
of the speed deviation $\delta$ from the equilibrium speed are plotted
within a region, where the stability is lost for more than half of
the samples $\boldsymbol{B}$, i.e., the median of $\delta$ exceeds
the initial perturbation of $2m/s$. (c) Heatmap for the average deviation
$\delta$ from the stationary speed $v_{\ast}=20m/s$. Higher (lighter)
values at a coordinate indicate that, on average, the equilibrium
is less stable, resp. more unstable, for the corresponding parameter
combination than at coordinates exhibiting lower values. The intervals
corresponding to (a) and (b) are indicated as well as to specific
values of $\Delta\hat{B}$ for the homogeneous system: (i) equilibrium
stability switch, (ii) regime switch from classical Gipps to tangential
calculus introduced in Section~\ref{sec:vsafe-for-b-ge-bhat}.}
\end{figure}

Figure~\ref{fig:stability-scan} shows a numerical stability experiment
for the proposed extension of the Gipps model, where the magnitude
of heterogeneity $\Delta B$ and the error $\Delta\hat{B}$ in the
estimation of the leader's braking capability is varied. For each
combination of the parameters we initialized the heterogeneous system
in the stationary state for $v_{\ast}=20m/s$. Then, we then perturbed
this state by setting the initial velocity of a single vehicle to
$v_{i}(0)=v_{\ast}-\tilde{v}$, with an initial perturbation $\tilde{v}=2m/s$,
and monitored the evolution of the maximal deviation $\delta=\left|v_{i}(t)-v_{\ast}\right|$
of all vehicles' speeds from $v_{\ast}$. If the equilibrium flow
was unstable, the perturbation would grow and the deviation could
eventually reach its maximal possible value at $v_{\ast}=20m/s$.
This maximal deviation could be observed if stop-and-go waves emerged,
where vehicles came to a total stop, or was imposed automatically
if collisions were detected. For each parameter point, we executed
several simulation runs of duration $t_{{\rm end}}=1000s.$ (Fig.~\ref{fig:stability-scan}(a),(b)),
resp. $t_{{\rm end}}=500s.$ (Fig.~\ref{fig:stability-scan}(c)),
with different realizations of $\boldsymbol{B}$. Figure~\ref{fig:stability-scan}(a)
and (b) show the destabilization regions in $\Delta\hat{B}$ for fixed
values $\Delta B=0.5$ and $\Delta B=2.0$, respectively. To give
an impression of the variability of the system's behavior across different
realizations of $\boldsymbol{B}$, we show several percentiles for
the observed distributions of the deviation $\delta$, where we only
considered times $t>T/2$ for the calculation of $\delta$. In general,
a value $\delta>\tilde{v}$ indicates an instability of the equilibrium
flow while $\delta<\tilde{v}$ indicates stability. Notably, the
case $\Delta B=2.0$ of larger heterogeneity exhibits a larger region
of mixed stability as the case $\Delta B=0.5$ and the stabilization
takes place earlier when varying $\Delta\hat{B}$ from negative values
towards the error-free case $\Delta\hat{B}=0$.

Figure~\ref{fig:stability-scan}(c) shows a heatmap, whose color
indicates the mean of the observed deviations $\delta$ for a parameter
region $\Delta B\in[0.0,2.0]$ and $\Delta\hat{B}\in[-0.5,0.0]$.
Each pixel represents a parameter combination and $50$ corresponding
simulation runs, which were initialized as described above and run
for a duration of $t_{{\rm end}}=500s$. Two red bars indicate the
one-dimensional parameter slices corresponding to (a) and (b). The
crosses at $\Delta B=0.0$ indicate the values of $\Delta\hat{B}$
for the homogeneous system, at which (i) the equilibrium flow of the
classical Gipps ensemble destabilizes ($\Delta\hat{B}\approx-0.142$,
cf. \cite[ Eqn. (5.13)]{wilson_analysis_2001}), and (ii), $\Delta\hat{B}\approx-0.388$,
beyond which the equilibrium regime is governed by the tangential
calculus, cf. condition (\ref{eq:cond-for-alpha2-in-equilibrium-2}).
In accordance with the analytic results presented in Section~\ref{sec:Stability-of-the-equilibrium}
the equilibrium is unstable for $\Delta\hat{B}<0.388$ and $\Delta B=0.0$.
It may be observed that for all values of $\Delta B$, increasing
the magnitude of the shift $\Delta\hat{B}$ acts destabilizing. Though
this effect is clearly ameliorated for larger variations $\Delta B$
of the maximal desired braking rates, indicating a stabilizing effect
attributable to the ensemble heterogeneity.

\section{Discussion\label{sec:Discussion}}

In the present work we introduced a novel, though canonic, approach
to resolve collisions occurring in the Gipps car-following model,
which was originally devised by the principle of choosing the maximal
safe acceleration for the next simulation speed. Although mostly omitted,
we assume that it is well known, that the safety principle fails in
heterogeneous ensembles of vehicles, where the desired braking rates
$B_{i}$ of the different vehicles differ to a sufficient degree.
We pinpointed the underlying reason for the collisions occuring in
that situation and proposed a solution, which follows the original
idea of travelling at a maximal safe speed. We kept the original form
of Gipps' considerations for choosing the safe speed by assuming a
three-partite continuation of a vehicle's trajectory into phases of
constant acceleration, constant speed, and constant deceleration {[}cf.
Section~\ref{sec:Gipps-calculus-reviewed}{]}, and we showed that
for some cases the maximal safe speed is not determined by the speed
obtained from the Gipps model, but by a calculation of a tangency
of the follower's distance $g(t)$ to the leader in the hypothesized
trajectory continuation, i.e., an occurrence of $g(t_{\parallel})=g^{\prime}(t_{\parallel})=0$
for some $t_{\parallel}>t$. We analyzed the stability of the equilibrium
flow in a regime, where the tangential calculus acts limiting on the
stationary speed and found that it is always unstable. This is due
to the requirement of a large underestimation $\Delta\hat{B}=\hat{B}_{i}-B_{i}$
of the braking rate by interacting vehicles to 'activate' the corresponding
regime.\textcolor{red}{{} }For heterogeneous ensembles, where the tangential
case becomes relevant even without a misestimation of the braking
rates, we conducted numerical stability experiments. Interestingly,
these indicate an positive relation for the robustness of the equilibrium
flow and the measure of heterogeneity. That is, stability is enhanced
by heterogeneity. 

One caveat for the proposed extension of the Gipps model is the singularity
in the speed-headway relation obtained from the new regime, see Fig.~\ref{fig:equilibrium-speed},
whose discontinuety does not meet the expectable form for this relation.
A possible approach to handle this flaw might be the introduction
of a speed dependence in the desired braking rate $B(v)$. Considering
this characteristic, one may be tempted to retreat to the simpler
strategy of resolving the collisions in the Gipps model by choosing
always the maximum of $B_{i}$ and $\hat{B}_{i+1}$ for the calculation
of the safe following speed as, for instance, exercised by the traffic
simulation SUMO \cite{SUMO2012}. However, given standard assumptions
this will always lead to a stable flow as instabilities require $\hat{B}_{i+1}<B_{i}$
\cite{wilson_analysis_2001}, and phenomena usually assumed to be
related to instabilities must then be imposed as additional model
components, see for instance the 'dawdling' in the Krauss model \cite{Krauss1997,Kraus1998}.

Keeping these remarks in mind is especially important when one is
concerned with simulations of safety aspects of traffic, such as simulated
safety surrogate measures, for instance {[}CITE Gettman{]}. In any
case, other possible ways to provoke unsafe or imperfect behavior
of the Gipps model are assuming negative values $\theta<0$ for the
reaction time buffer $\theta$, or values $\tau<dt$ for the assumed
reaction time $\tau$, which both may be interpreted as a driver's
misestimation of the own reactive capabilities. Further, errors may
be imposed on the input or output \cite{Krauss1997,Kraus1998,Eissfeldt2003}
of the model to achieve safety relevant deviations or plainly an increased
realism for the model.

\bibliographystyle{plain}
\bibliography{GippsExtended}

\end{document}